\documentclass[11pt,oneside,letterpaper]{article}
\usepackage{amssymb}
\usepackage{amsmath}
\usepackage[dvips]{graphicx}
\usepackage{setspace}
\usepackage{amsfonts}
\usepackage{fancyhdr}
\usepackage{xcolor}
\usepackage{graphicx}
\usepackage{rotating}
\usepackage{comment}
\usepackage{color}
\usepackage{cite}
\usepackage{braket}
\usepackage{caption}
\usepackage[utf8]{inputenc}
\usepackage[c]{esvect}
\usepackage{mathabx}
\usepackage{framed}

\definecolor{darkgreen}{rgb}{0,0.5,0}
\definecolor{darkblue}{rgb}{0,0,0.6}
\definecolor{purple}{rgb}{0.4,.2,0.7}

\renewcommand{\ket}{\rangle}
\renewcommand{\bra}{\langle}
\newcommand{\be}{\begin{equation}}
\newcommand{\ee}{\end{equation}}

\usepackage{pdfsync}
\makeatletter
\newcommand*{\defeq}{\mathrel{\rlap{%
                     \raisebox{0.3ex}{$\m@th\cdot$}}%
                     \raisebox{-0.3ex}{$\m@th\cdot$}}%
                     =} 
\makeatother

\def\be{\begin{eqnarray}}
\def\ee{\end{eqnarray}}

\newcommand{\bea}{\begin{eqnarray}}
\newcommand{\eea}{\end{eqnarray}}
\def\ben{\begin{equation}}
\def\een{\end{equation}}

     \let\r=v

\def\be{\begin{equation}}
\def\ee{\end{equation}}
\def\ba{\begin{array}}
\def\ea{\end{array}}

\def\ba#1\ea{\begin{align}#1\end{align}}
\def\bs#1\es{\begin{split}#1\end{split}}

\newcommand{\At}{A_\times}

\allowdisplaybreaks  % allow page breaks in displayed eqs

\numberwithin{equation}{section}
\def \be {\begin{equation}}
\def \ee {\end{equation}}
	
\def \JM#1 {{\color{blue}  JM: #1 }}
\def \AAl#1 {{\color{red}  AA: #1 }}

\begin{document}
\onehalfspacing

\begin{center}

~
\vskip5mm

{\LARGE  {
Subadditive Average Distances and 
Quantum Promptness}\\
\ \\
}

Federico Piazza$^1$ and  Andrew J. Tolley$^2$

\vskip5mm
{\it $^1$Aix Marseille Univ, Universit\'{e} de Toulon, CNRS, CPT, Marseille, France. \\
$^2$Theoretical Physics, Blackett Laboratory, Imperial College, London, SW7 2AZ, U.K.} 

\vskip5mm

{\tt piazza@cpt.univ-mrs.fr, a.tolley@imperial.ac.uk}

\end{center}

\vspace{4mm}

\begin{abstract}
\noindent

A central property of a classical geometry is that the geodesic distance between two events is \emph{additive}. When considering quantum fluctuations in the metric or a quantum or statistical superposition of different spacetimes, additivity is generically lost at the level of expectation values. In the presence of a  superposition of metrics, distances can be made diffeomorphism invariant by considering the frame of a family of free-falling observers or a pressureless fluid, provided we work at sufficiently low energies. 
We propose to use the average squared distance between two events $\bra d^2(x,y)\ket$ as a proxy for understanding the effective quantum (or statistical) geometry and the emergent causal relations among such observers.  At each point, the average squared distance $\bra d^2(x,y)\ket$ defines an average metric tensor. However, due to non-additivity, $\bra d^2(x,y)\ket$ is not the (squared) geodesic distance associated with it. 
We show that departures from additivity can be conveniently captured by a bi-local quantity $C(x,y)$.  Violations of additivity build up with the mutual separation between $x$ and $y$ and can  correspond to $C<0$ (subadditive) or $C>0$ (superadditive). We show that average Euclidean distances are always subadditive: they satisfy the triangle inequality but generally fail to saturate it. In Lorentzian signature there is no definite result about the sign of $C$, most physical examples give $C<0$ but there exist counterexamples.  The causality induced by subadditive Lorentzian distances is unorthodox but not pathological. Superadditivity violates the transitivity of causal relations. On these bases, we argue that subadditive distances are the expected outcome of dynamical evolution, if relatively generic physical initial conditions are considered.

 \end{abstract}
\vspace{.2in}
\vspace{.3in}

\pagebreak
\pagestyle{plain}

\setcounter{tocdepth}{2}
{}
\vfill

\ \vspace{-2cm}
\newcommand{\deq}{{\overrightarrow {\Delta x}}^{\, 2}}
\renewcommand{\baselinestretch}{1}\small
\tableofcontents
\renewcommand{\baselinestretch}{1.15}\normalsize

\section{Introduction}
 
On a classical spacetime, the geodesic distance between two points $x$ and $y$ is additive in the sense that for any intermediate point $z$ on the geodesic,
\be\label{addd}
d(x,y) = d(x,z)+d(z,y) \, .
\ee
This follows from the fact that the distance can be written as a single integral of a local function of the trajectory $x^{\mu}(\lambda)$ 
\be
d(x,y)= \int_0^1 d \lambda \sqrt{g_{\mu\nu}(x(\lambda)) \frac{d x^{\mu}}{d \lambda}  \frac{d x^{\nu}}{d \lambda}} \, .
\ee
For finite separation the distance depends non-linearly on the metric for fixed end points. As such when quantum fluctuations of the metric are included, or when a state is considered which contains a superposition of (semi-)classical metrics, this straightforward property of additivity is generically lost at the level of expectation values.\footnote{For example, in Euclidean signature, when computing $\langle d(x,y) \rangle$ we sum over all paths which begin at $x$ and end at $y$. This a different set of paths than for $d(x,z)$ and so the direct average of \eqref{addd} does not hold $ \langle d(x,y)\rangle  \neq \langle d(x,z)\rangle+\langle d(z,y)\rangle$ .}

To give a gauge invariant meaning to the geodesic distance $d(x,y)$ we must first give a physical anchor to the end-points $x$ and $y$. Defining observables in quantum gravity is notoriously subtle because of gauge invariance. Strictly speaking no local gauge invariant observables exist, in the face of which many authors resort to considering only S-matrix elements as observables, or boundary correlation functions in the case of asymptotically AdS spacetimes. To define even an approximately local observable, one needs a prescription for identifying \emph{which point corresponds to which} within the different geometries that compose the statistical ensemble. When the fluctuations of the metric are large this is highly non-trivial. In asymptotically AdS space one way of dealing with this is to use geodesics to anchor bulk points to the boundary and dress the corresponding operators (e.g.~\cite{Giddings:2018umg} and references therein). The resulting operators are gauge invariant but mildly non-local. 

Geodesics can be used to define an event regardless of the asymptotic structure of the spacetime, as long as we remain within the domain of validity of some low energy theory. Consider a {\it dilute gas} of macroscopic observers free-falling along geodesic worldlines. At suitably low energies, the macroscopic objects can be consistently treated as background non-dynamical sources interacting with gravitons, as in the formalism of Goldberger and Rothstein for binary sources~\cite{Goldberger:2004jt}. These worldlines make a nice reference system of ``observers" that can enquire about the spacetime around them, similarly to what we do when we look at the sky from our planet. Provided two such observers can synchronize their clocks and their relative positions well in the past,  an event is \emph{defined}  by the initial position of the worldline and by the value of the proper time along it. As long as the observers can continue to be described by a worldline effective theory, meaning in practice that we work at sufficiently low energies, then the wordlines serve to act as a coordinate system for the future evolution. As discussed in Sec.~\ref{sec_observers}, in the continuum limit the dilute gas of observers can be interpreted as a pressureless (or non-relativistic) fluid, and the preferred coordinate system will be {\it unitary gauge} for the fluid.\footnote{This approach differs from that of Geheniau and Debever and DeWitt \cite{DeWitt,Marolf:2015jha} in that rather than introducing $4$ scalar fields $Z^{\alpha}$ to specify a coordinate system, we use 3 scalars $X^I$ which define the pressureless fluid, with the time fixed by the integral along the worldline at a given $x^I$.}

To illustrate this, let us use an \emph{ad-hoc} but highly informative example, treated in detail in App.~\ref{grav-waves}.   Consider a gravitational plane wave traversing Minkowski space with wavefront at coordinates $x^3 = x^0$. For every given intensity and polarization we can associate a coherent state for the graviton $|\psi_i\rangle$ to this classical wave. As long as the quantum state of the system is described by a single coherent state, we may interpret the dynamics in terms of a single classical background geometry that is equivalent to Minkowski for $x^0<x^3$ and looks like a plane wave for $x^0>x^3$. More generally though we can also consider coherent superpositions of these classical states,\footnote{Similar quantum superpositions of geometries are considered in \cite{Christodoulou:2018cmk,Belenchia:2018szb,Foo:2022dnz}.}
\begin{equation} \label{quantum-ensemble}
|\psi\rangle = c_1 |\psi_1 \rangle + c_2 | \psi_2\rangle + \dots \ . 
\end{equation}
By construction, the state $|\psi\rangle$ is identical to Minkowski vacuum for $x^0<x^3$ but for $x^0>x^3$ can no longer be interpreted in terms of a single classical geometry. The fact that the asymptotic geometry for $x^0<x^3$ is Minkowski means that, at early times, each triplet of Minkowski spatial coordinates $\vec x$  labels a potential geodesic observer, whose proper time is initally at least simply the Minkowski time coordinate $x^0$. After $x^0 =x^3$, the dilute gas of observers, still free-falling, are gently stirred around by the gravitational waves superposition. 
In most coordinate systems (e.g. in ``Brinkmann coordinates", see App.~\ref{grav-waves}) they follow different trajectories on each different classical representative $|\psi_i\rangle$ of the quantum ensemble~\eqref{quantum-ensemble}, giving a Schr\"odinger's cat-like state for which the position of the observer appears to be indefinite. It is convenient, however, to take the observers' point of view and use the labels $\vec x$ and their proper time $x^0$ to \emph{define} an event in the quantum spacetime $|\psi\rangle$. For plane waves, this corresponds to using the so-called ``Rosen coordinates". As discussed in Sec.~\ref{sec_observers}, this is also the \emph{unitary gauge} for a fluid of non-relativistic particles. 
 
In this setup one  can calculate the outcomes of various gedanken experiments.  
For example, if observer $\vec x$ sends a photon at time $x^0$,  what is the probability $P$ that observer $\vec y$ receives the photon at time $y^0$? In the geometrical optics approximation, $P(y^0)$ can be calculated  by summing over the different null geodesics reaching $\vec y$ in the different geometries of the ensemble~\eqref{quantum-ensemble}. From $P(y^0)$ one can obtain e.g. the average time of arrival $\langle y^0\rangle$ and its variance $\Delta y^0$ (see figure). The fluctuations in the arrival time are determined by the fluctuations in the geometry, but not in the fluctuations of the observer wordlines which are fixed in this coordinate system.

\begin{figure}[h]
%\vspace{-1cm}
\begin{center}
   \includegraphics[width=10cm]{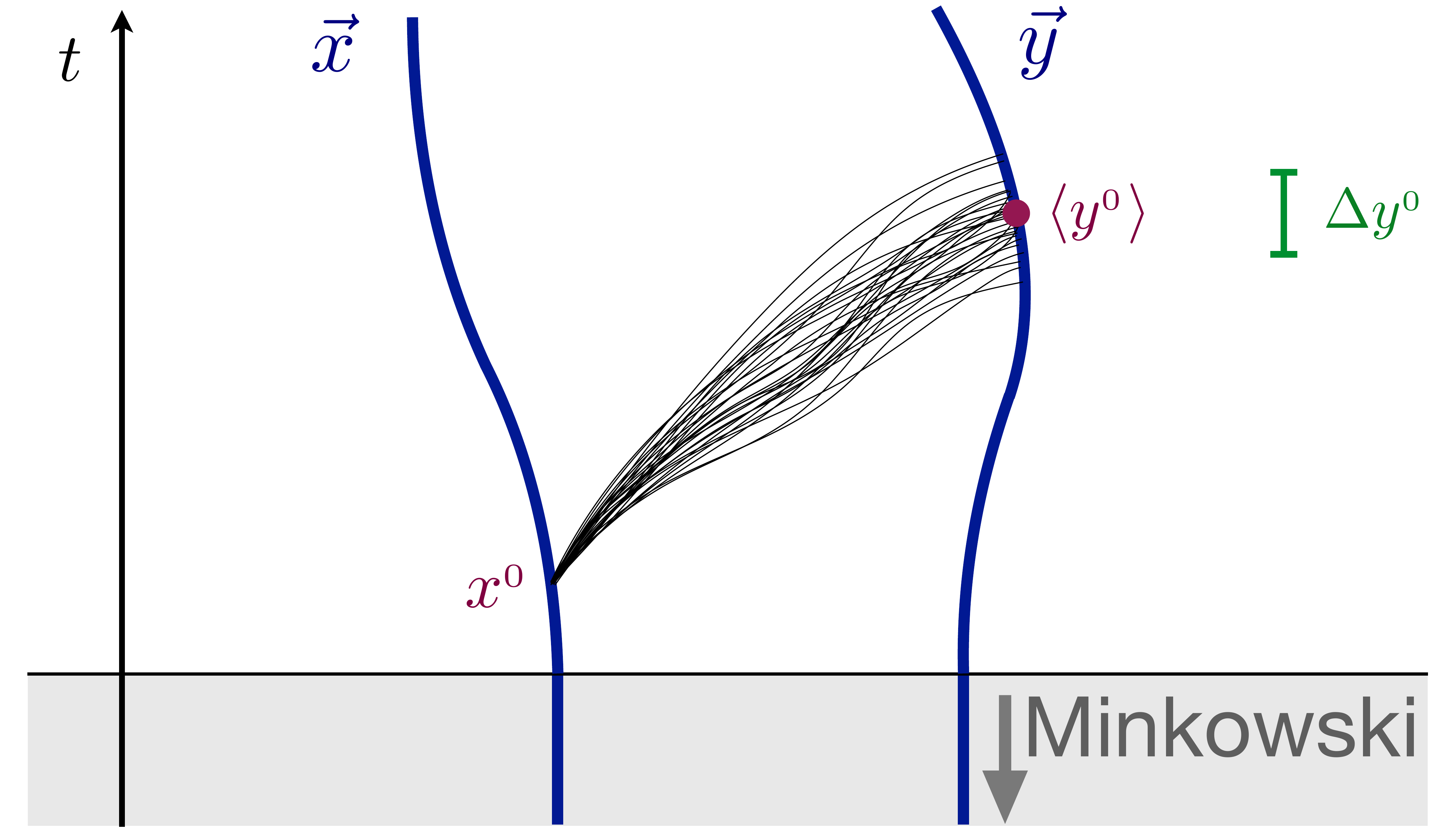}
  \end{center}\vspace{-.8cm}
   \end{figure}
A related, more involved, calculation would consist of dressing the local operators of some field theory by using these free-falling observers, as sketched in~\cite{Marolf:2015jha}. These dressed operators contain non local combinations of the metric field. On the state~\eqref{quantum-ensemble}, their commutators will have expectation values that do not sharply vanish outside any specific light cone, as generally expected in the presence of large quantum fluctuations of the gravitational field. 

Having fixed our coordinate system, we would like to better understand the generic features and implications of having a classical or quantum superposition of metrics. The fluctuations in the spacetime geometry lead to fluctuations in the gauge invariant distance between any two physical events as exemplified by the above gedanken experiment. One diagnostic that defines the resulting quantum geometry which shall be used throughout this paper, is the \emph{squared distance operator}\footnote{The Euclidean-signature version of this operator was studied in~\cite{Piazza:2021ojr}} between two locations, 
$d^2(x,y)$ which is well defined at least locally for $x$ and $y$ sufficiently close\footnote{The squared distance operator is (up to a factor) the same as Synge's world function $\sigma(x,y)=d^2(x,y)/2$ which plays a central role in his remarkable treatise on General Relativity \cite{Synge:1960ueh} and is a crucial ingredient in the Hadamard construction of Greens functions on curved spacetime \cite{DeWitt:1960fc,Harte:2012uw}.}. Its expectation value on each coherent state $|\psi_i\rangle$  is just the geodesic distance between $x$ and $y$ on the corresponding classical geometry. By contrast, when we consider superpositions of coherent states, the average of $d^2(x,y)$ will no longer correspond to the squared distance computed with some effective average metric.

For two nearby events, one can express $d^2(x,y)$ in terms of the metric operator and its derivatives in a coordinate expansion around  $x$ or $y$ (see Sec.~\ref{sec_normal}). This allows us to write the mildy non-local operator $d^2(x,y)$ in terms of an expansion in local operators which are more straightforwardly calculated. There are many good reasons for considering averages of the distances \emph{squared} rather than the distance $d(x,y)$ itself. First and foremost, they have a real spectrum and it makes sense to take averages of them---(Lorentzian) distances turn imaginary at timelike separation.  
In what follows, our goal is to explore the generic properties of the average squared distance, and in particular to understand to what extent we may regard it as an effective metric space. For example, we can ask at which location along the $\vec y$ worldline does $\langle d^2(x,y)\rangle=0$, for fixed $x$. The set of events at null average distance from $x$ traces an ``average light cone" that intersects $\vec y$ at some proper time $y^0$. For a quantum state that may be interpreted as a superposition of two distinct coherent (classical) states, this average light cone will be distinct from the light cones of either of the two classical geometries.
 
 In this paper we use the squared distance operator as a unique \emph{bona fide} proxy for all types of experiments/observables involving causality and more generically we regard the average $\langle d^2(x,y)\rangle$ as a proxy for the influence on the dynamics of matter on the superposition of geometries. We discuss in Sec.~\eqref{sec_meaning} how this is related to correlation functions of local operators, in particular the retarded propagator which encodes causal evolution. In the context of the above gedanken experiment, we expect that the average time of arrival $\langle y^0 \rangle$ of the photon be well approximated by the value of $y^0$ where $\langle d^2(x,y)\rangle=0$, although the two events need not strictly coincide and it is in this sense that our observable is only a proxy.

 One might expect to recover from this analysis a classical spacetime ``perturbed" by quantum fluctuations. Or, that the uncertainty of the distance operator could induce occasional violations of causality on top of an otherwise classical light cone structure. We find instead that \emph{the very expectation value of the distance behaves anomalously} and portraits an ``average causal structure" that has no classical analogue.  In particular, we find that average distances in quantum gravity are \emph{non-additive}, as opposed to the standard geodesic distances of a classical spacetime. 

\section{Summary}
\subsection{Additivity and lack thereof}

Consider a spacetime metric $g_{\mu \nu}$ and the associated geodesic distance $d(x, y)$.  The latter must satisfy a non-trivial property which we dub \emph{additivity}.  Here we just sketch what we mean by additivity in the case of null separation, that of most direct relevance for causality. A more throughout  analysis is provided in Sec.~\ref{additivity}.

Consider two points $x$ and $z$ at null distance, $d(x,z) =0$, and the following two equations for the unknown intermediate point $y$, 
\begin{equation} \label{nullness}
 d(x,y) =0 ; \qquad d(y,z)= 0. 
  \end{equation}
Two conditions potentially define a codimension-2 submanifold. However, here the solution is degenerate. The points $y$ satisfying~\eqref{nullness} make a one-dimensional manifold, namely, the null geodesic connecting $x$ and $z$. This is what defines additivity in the case of null distance. Clearly, we are assuming sufficiently ``standard" conditions---for example,  the absence of conjugate points \cite{Synge:1960ueh}. One way of seeing this degeneracy is to realize that  null-separation is the limiting case between timelike  and spacelike separation. 
If we slightly move $z$ to make it timelike with respect to $x$, the forward light cone from $x$  intersects the backward light cone from $z$ on a codimension-two surface (in Minkowski the intersection is a metric two-sphere). If instead $x$ and $z$ are at spacelike separation the two light cones do not intersect and~\eqref{nullness} has no real solution.

Because it corresponds to a degenerate case, additivity is a fragile property, which is lost at the level of \emph{expectation values}. Once we manage to calculate $\bra d^2(x,y) \ket$ on some gravitational state $|\psi\ket$ we can define a (pseudo-)Riemannian structure \emph{locally}, in the vicinity of each point. If  
$\bra d^2(x,y) \ket$ is sufficiently smooth one can define an \emph{average metric tensor},
 \begin{equation} \label{metricave}
\bar g_{\mu \nu} (x) \ \equiv \ -\frac12 \, \lim_{y\rightarrow x} \, \frac{\partial}{\partial x^\mu} \frac{\partial}{\partial y^\nu} \bra d^2(x,y)\ket \, .
\end{equation}
It is also convenient to introduce the following quantity,
\begin{equation}
\bar d(x,y) \ \equiv \ \sqrt{\bra d^2(x,y)\ket},
\end{equation}
and (mis)call it \emph{average distance} for short. The lack of additivity manifests itself in the fact that although we can always associate a metric tensor to $\bar d(x,y)$ through~\eqref{metricave}, the reverse is not true. The bi-scalar $\bar d(x,y)$, in general, is \emph{not} the geodesic distance of any metric and we need more information than $\bar g_{\mu \nu}$ to reconstruct it.

Having constructed the average distance, we can propose again the problem encapsulated by eqs.~\eqref{nullness}. Broadly speaking, two things can happen: It could be that  $x$ and $z$ are at null-separation $\bar d(x, z)=0$, 
and yet 
\begin{equation} \label{nullness2}
 \bar d(x,y) =0 \, , \qquad \bar d(y,z)= 0 \, ,
  \end{equation}
has no real solution. In this case we call $\bar d$ \emph{subadditive}. If instead the solutions $y$ span a co-dimension two manifold we say that $\bar d$ is \emph{superadditive}. 

To gain some intuition, one can find these types of non-additive behaviors in \emph{chordal distances}. Consider a pseudo-Riemannian manifold embedded in spaces with one additional space dimension. If the lower dimensional manifold is bent like a cylinder along a space direction, then the light rays in the ambient space will arrive sooner than those constrained along the cylinder. For this type of embedding the chordal distance is  \emph{subadditive}---it is easy to check that~\eqref{nullness2} has no solution. Wrapping along the time direction has the opposite effect and gives a \emph{superadditive} distance. In this analogy, $\bar g_{\mu \nu} (x)$ is the intrinsic metric of the embedded manifold. But in order to calculate chordal distances one needs additional  information, e.g., about the extrinsic curvature of the embedding. 

\subsection{Measuring non-additivity}

There is a useful quantity that one can calculate to measure the non-additivity of $ \bar d(x,y)$. As for $ \bar d(x,y)$ itself, such a quantity is a bi-scalar, which we introduce in Sec.~\ref{additivity}, 
\begin{equation} \label{crucial-intro}
C(x,y) \equiv \frac14\,  \frac{\partial \,  \bra d^2(x,y)\ket}{\partial y^\mu} \frac{\partial \, \bra d^2(x,y)\ket}{\partial y^\nu} \ \bar g^{\mu \nu}(y) - \bra d^2(x,y)\ket\, .
\end{equation}
In the above $\bar g^{\mu \nu}$ is the inverse of $\bar g_{\mu \nu}$.  When calculated for a standard geodesic distance $C$  is exactly zero. This follows from direct calculation, but is more easily understood by considering the physical example for which $d(x,y)$ denotes the distance along the world-line of a massive particle of mass $m$. Its action or Hamilton-Jacobi function evaluated on a trajectory from $x$ to $y$ is determined by the proper time $S(x,y)= - m \int d \tau \sqrt{-g_{\mu\nu} \dot x^{\mu}(\tau)  \dot x^{\nu}(\tau)}= - m \sqrt{-d^2(x,y)}$. At the same time, the action will satisfy the following Hamilton-Jacobi equation
\be \label{Hamilton-Jacobi}
g^{\mu\nu}\partial_{\mu} S\partial_{\nu}S+m^2=0
\ee
by virtue of worldline reparameterization invariance. Substituting in $S(x,y)= - m \sqrt{-d^2(x,y)}$ we recognize that \eqref{Hamilton-Jacobi} is equivalent to $C(x,y)=0$. The same argument can be applied to spacelike geodesics by replacing $m \rightarrow i m$. When we consider averages over distances for classical or quantum superpositions of different geometries this relation no longer holds. Subadditivity (superadditivity) corresponds to $C<0$ ($C>0$). \\

It is useful to express the geodesic distance explicitly in some coordinate system.  In Sec.~\ref{sec_normal}  the distance is Taylor expanded around one of the two extremes, under the hypothesis of staying within its normal neighborhood. It is convenient to set the origin of the coordinates on such a point. The result reads
\begin{align} \label{dsquare2-intro}
d^2(0,x) =& \ g_{\mu \nu}x^\mu x^\nu + \frac12 g_{\mu \nu, \rho} \, x^\mu x^\nu x^\rho 
 - \frac{1}{12}\left(g_{\alpha \beta}\Gamma^\alpha_{\mu \nu} \Gamma^\beta_{\rho \sigma} - 2 g_{\mu \nu, \rho \sigma}\right) x^\mu x^\nu x^\rho x^\sigma + {\cal O}(x^5)\, .
\end{align}
where the metric and its derivatives are calculated at $x^\mu = 0$. 
If we are in the presence of a statistical ensemble of metrics, $\bra d^2(0,x) \ket$ is obtained by taking the average of each term of this expansion.  
If the coordinates $x^\mu$ are ``physical", \emph{i.e.} if they are independently defined as a dynamical reference system, the averaging hits only the metric and its derivatives.\footnote{If the gauge is instead fixed by requiring that the metric has some given form, e.g. as in the case of Riemann normal coordinates, the coordinates themselves become operators, and one should put them inside the average.} This is the case for the  system of macroscopic free-falling observers sketched in the introduction. 
To lowest order, $d^2(0,x) = \bra g_{\mu \nu} (0) \ket x^\mu x^\nu$, which, through~\eqref{metricave}, means that at each point we can in fact identify $\bar g_{\mu \nu}(x) = \bra g_{\mu \nu} (x)\ket$. 

We are finally able to evaluate~\eqref{crucial-intro}, to which only the quadratic terms in~\eqref{dsquare2-intro} contribute,
\begin{equation}\label{c-expansion}
C(0,x) = \frac14\left(\bar g^{\alpha \beta}\langle \Gamma_{\alpha \mu \nu} \rangle \langle \Gamma_{\beta \rho \sigma} \rangle - \langle g_{\alpha \beta}\Gamma^\alpha_{\mu \nu} \Gamma^\beta_{\rho \sigma}\rangle\right) x^\mu x^\nu x^\rho x^\sigma + {\cal O}(x^5)\, ,
\end{equation}
where $\Gamma_{\alpha \mu \nu} \equiv g_{\alpha \beta} \Gamma^\beta_{ \mu \nu}$.
Notice that $C$ starts at order ${\cal O}(x^4)$ and, as announced, it vanishes when averages are neglected, i.e., on a classical spacetime. Once again, the analogy with chordal distances is pertinent because also for them $C$ starts no sooner than at ${\cal O}(x^4)$. 
As qualitatively anticipated in~\cite{Piazza:2021ojr}, non-additivity is a property that builds up at large separation.

\subsection{The character of average distances, the size of the effect and causality}

One main result of this paper is that in Euclidean signature average distances are always \emph{subadditive}. As detailed in Sec.~\ref{additivity} this means that average Euclidean distances are still \emph{distances} in the sense of \emph{metric spaces}: they satisfy the triangle inequality, although they generally fail to saturate it. We prove this result at the leading order in a coordinate expansion in Sec.~\ref{subsec-euc} by showing that the quantity in~\eqref{c-expansion} is semi-negative definite. For classical statistical ensembles of Euclidean-signature metrics we show evidence of subadditivity beyond leading order in Appendixes~\ref{app-resum} and~\ref{app-triangle}.

In Lorentzian signature there is no sharp result about the character of average distances. However, also in this case subadditivity seems more generic than superadditivity. We calculate~\eqref{c-expansion} perturbatively in a thermal bath of gravitons (App.~\ref{app-thermal}). Symmetry considerations learned from this example allow to conclude that  average distances are always subadditive for small gravitational perturbations around an homogeneous background (see Sec.~\ref{subec-lorentzian}).  Naively, subadditivity seems to follow from the fact that ``space wins over time" as number of dimensions. Another (non-perturbative) example of subadditivity is made by general superpositions of plane waves solutions (Sec.~\ref{sec_planewaves}).
There exists, however, examples where average distances are superadditive, like the superposition of FRW metrics with sufficiently negative equation of state discussed in Sec.~\ref{FRW}.

The violation of additivity is relatively tiny in perturbation theory and is necessarily suppressed by the Planck scale squared. However we expect it to be possibly enhanced in non-perturbative situations. In field theory, those are the situations where, broadly speaking, Euclidean saddles of the path integral interpolate different classical vacua. In those cases the vacuum of these theories is a coherent superposition of macroscopically different configurations. It is tempting to wonder if something similar could be at play during black hole evaporation, where non-trivial Euclidean saddles connecting different boundaries account for the decrease of entropy in the late-Hawking radiation~\cite{Almheiri:2020cfm}. Subadditivity could give a handle to make sense of the information leaking to infinity from the point of view of the observers living inside the spacetime.  
The reason is that non-additive distances display a light cone structure that the average metric $\bra g_{\mu \nu}\ket$ can only approximate \emph{locally}.  In a classical spacetime, in the presence of strong lensing events, a light ray is \emph{prompt} if it arrives before the others emitted from the same source~\cite{Witten:2019qhl}. For subadditive distances,  the classical causal structure of the average metric $\bra g_{\mu \nu}\ket$ \emph{ overestimates} the times of arrival of a light ray. In other words, quantum gravity seems to contain a mechanism to make photons \emph{prompt} with respect to the classical expectation. 
This is all discussed at length in Sec.~\ref{qp}.

\section{Effective Field theory of a fluid of  observers} \label{sec_observers}

The physical coordinate system of geodesic observers can be formally introduced as a non-relativistic fluid coupled to gravity. In the limit of zero pressure, the elements of a fluid follow geodesics. There is a low energy description for this system~\cite{Dubovsky:2005xd,Endlich:2010hf}, corresponding to the action 
\begin{equation} \label{2}
S  \ = \  \frac{1}{16 \pi G}\int d^4 x \sqrt{-g} R \   - \ \mu^4 \int d^4 x \sqrt{-g} \sqrt{\det B^{IJ}}\  + \ S_m[\Phi] \ + \ \dots\, ,
\end{equation}
  where $\mu$ is a coupling with dimensions of mass, $ \Phi$ denotes all other matter fields and 
  \begin{equation}
B^{IJ} = g^{\mu \nu} \partial_\mu X^I \partial_\nu X^J\, .
\end{equation}
So a fluid is described in terms of three scalar fields  $X^I(x)$, $I = 1,2,3$. Each triplet is associated to a fluid element, or to an observer. There is clearly a continuum of them in this formalism.  The use of dynamical reference systems of this type in quantum gravity has also been discussed e.g. in~\cite{Rovelli:1990ph,Brown:1994py,Marolf:2015jha,Piazza:2021ojr}. 

The cutoff of this theory is $\mu$ and is taken to be much lower than that of gravity $\mu \ll M_{\rm Planck}$ to ensure that the backreaction of the fluid maybe negligible. Our macroscopic observers are stable objects that cannot annihilate into other fields, pair create, nor scatter among each other. Within these low-energy restrictions, the system can be considered at the full quantum mechanical level. The state is specified by a wavefunction of the fields, 
\begin{equation} 
\Psi = \Psi[h_{ij} (x^k), \, \Phi(x^k), \, X^I(x^k)]\, .
\end{equation}
As argued in the introduction, it is particularly convenient to take the ``point of view of the observers" which corresponds here to going into \emph{unitary gauge}. In other words, we directly chose the fields $X^I$ as the spatial coordinates, $X^I = x^i$. In this gauge the state only depends on the remaining fields, 
\begin{equation} 
\Psi = \Psi[h_{ij} (X^i), \, \Phi(X^i)]\, .
\end{equation}
   
There are interesting well known generalizations of~\eqref{2} that describe perfect fluids with arbitrary equations of state or solids~\cite{Dubovsky:2005xd}. One might wonder what is special about non-relativistic fluids. The point is that the comoving volume elements of a non-relativistic fluid are free falling, as we explicitly verify in App.~\ref{app-fluid}. We argue that this makes a better reference system. The dynamics of more general fluids and solids is dictated also by internal forces (pressure or anisotropic stresses). 
For each classical metric composing some quantum ensemble such internal stresses are different. Effectively, we would be dealing with a reference system that behaves differently according to the state of the metric field that we aim to probe.

Another nice quality that we demand from our observers is that of being hypersurface orthogonal on each classical metric composing the quantum ensemble. All we need to ask is that there is some given ``initial" hypersurface $\Sigma$ whose gradient is parallel to the four-velocities of the observers.\footnote{If one considers small perturbations around Minkowski or FLRW space this is not a restriction. In more general cases there could be issues related to global property of the spacetime.} On $\Sigma$ we can synchronize the proper times of the observers and the time coordinate can be defined as their proper time from there on.  The dynamics of a non-relativistic fluid guarantees that, if the unitary gauge shifts $g_{0i}$ vanish on $\Sigma$, they will remain vanishing along the evolution. This means that the hypersurface orthogonality is preserved in time, once it is postulated as an initial condition. This property (shown in App.~\ref{app-fluid}) is simply a rephrasing of Kelvin's circulation theorem, in that vorticity, if initially vanishing, cannot be created during evolution.

In summary, in unitary gauge for these observers, the metric has the form   
   \begin{equation}
g_{\mu \nu} dx^\mu dx^\nu = -dt^2 + \gamma_{ij} dx^i dx^j\, ,
\end{equation}
where from now on we understand lower case $x^i$ to be the coordinates of the fluid in unitary gauge.\footnote{From now, the notation will be opposite to that in~\cite{Piazza:2021ojr}: lower case letters are the physical coordinates related to the observers while all other coordinates (e.g. Riemann normal coordinates) will be indicated with capital letters.}

\section{The geometry of non-additive distances} \label{additivity}

Assume that a real bi-scalar  $d^2(x,y)$ is defined on a manifold.\footnote{In this section $d$ denotes a general, sufficiently smooth distance (additive or non-additive) defined on a manifold. We are not concerned, at present, with the global aspects of non-additive distances. The topology can be assumed to be $\mathbb{R}^3$ or $\mathbb{R}^4$  }  It must be \emph{symmetric}, $d^2(x,y) = d^2(y,x)$ 
and sufficiently smooth so that it defines a metric tensor at any point,
\begin{equation} \label{metrictensor}
g_{\mu \nu} (x) \ \equiv \ -\frac12 \, \lim_{y\rightarrow x} \, \frac{\partial}{\partial x^\mu} \frac{\partial}{\partial y^\nu} d^2(x,y)\, .
\end{equation}
We demand that $g$ be finite, nowhere degenerate and  smooth on the manifold. 
If $d^2(x,y)$ is positive definite then we further demand that $d^2(x,y)=0$ if and only if $x=y$. In this case $g$ has positive eigenvalues (\emph{Euclidean signature}).  \emph{Lorentzian signature} is instead characterized by $g$ having one negative and three positive eigenvalues. In this case we expect $d^2(x,y)$ to be either positive or negative. Its square root, $d(x,y)$, either positive real or positive imaginary. 

In standard classical spaces of either signatures, the intuitive idea that \emph{distances add up}  is clearly rooted in the fact that lengths are the  integrals of a differential line element. However, a geodesic distance is obtained by extremising such a length over all possible paths. And clearly, in more than one dimension, $d(x,y) + d(y,z) \neq d(x,z)$ in general.
So, in which precise sense geodesic distances are \emph{additive}? 
In order to characterize additivity, and lack thereof, here we propose to study the solutions of the \emph{third point problem}. 
\begin{framed}
\begin{quote}{\bf Third Point Problem (TPP).} Given two distinct points of coordinates $x$ and $z$ arbitrarily close to each other and a real positive number $R$,
\begin{equation}
0\leq R \leq |d(x, z)|\, , 
\end{equation}
Find a \emph{third point} $y$ such that 

If $d^2(x, z) \geq 0$,
\begin{align}
d(y, z) &= R \, ,\label{tpp1}\\
d(x, y) & = d(x, z) - R\, .\label{tpp2}
\end{align} 

If $d^2(x, z) < 0$,
\begin{align}
d(y, z) &= i R \, , \label{tpp3}\\
d(x, y) & = d(x, z) - i R\, .\label{tpp4}
\end{align} 

\end{quote}
\end{framed}
We start by discussing the case of Euclidean signature.
\subsection{Euclidean signature}  
We have already introduced a certain number of requirements on $d(x,y)$. The first, holding in any signature, is symmetry, 
\begin{enumerate}
\item[$i)$] $d(x,y) = d(y,x)$  ,
\end{enumerate}
the second is that
\begin{enumerate}
\item[$ii)$] $d(x,y) = 0$\qquad if and only if \qquad $x=y$ .
\end{enumerate}
Geodesic Riemannian distances define a \emph{metric-space} so an additional property that one might want to consider is the \emph{triangle inequality}
\begin{enumerate}
\item[$iii)$] $d(x,z) \leq d(x,y) + d(y,z)$.
\end{enumerate}
\subsubsection{Additivity}
Geodesic Riemannian distances are additive in the sense that one can always find a point $y$ that \emph{saturates} the triangle inequality. In particular, any point $y$ on the geodesic connecting $x$ and $z$ does. If we also ask that $y$ is at some given distance $R$ from $z$ as required by the TPP, then the solution is unique. So \emph{additivity in Euclidean signature is defined by TPP having one and only one solution.} A natural objection is that if $x$ and $z$ are conjugate points there could be more than one geodesic connecting them with the same length. In this case there is more than one point $y$ saturating $iii)$. But this does not happen if $x$ and $z$ are in the normal neighborhood of each other. This is why we formulated the TPP for points $x$ and $z$ that are \emph{arbitrarily close}. 

The obvious example of additive distance is the standard distance in \emph{Euclidean space},
\begin{equation} \label{ex-euc}
d^2(x_1,x_2) = (x_1^i - x_2^i) (x_1^j - x_2^j) \delta_{ij}\, \equiv \deq.
\end{equation}

\subsubsection{Subadditivity}
It is known that if a distance satisfies all the metric-space properties above, then a strictly convex function of it, $f(d(x,z))$, still defines a metric-space, provided that $f(0) = 0$ and $f'(0)=1$. In other words, the new distance still satisfies the triangle inequality, but fails to saturate it. In this case the TPP has no solution. \emph{So subadditivity in Euclidean signature is defined by TPP having no real solution.} A simple example of a subadditive distance is something that goes like 
\begin{equation} \label{ex-2}
d^2(x_1,x_2) = \deq \left[1 - \epsilon \, \deq \right]\, ,  
\end{equation}
with $\epsilon>0$. It is easy to verify that chord distances of embedded manifolds are subadditive. 

\subsubsection{Superadditivity}
This is not a very interesting case and can be obtained by taking a concave function of an additive distance, for example, 
\begin{equation} \label{ex-3}
d^2(x_1,x_2) = \deq \left[1 + \epsilon \, \deq\right]\, .  
\end{equation}
The above is not a distance in the metric-space sense because the triangle inequality $iii)$ is failed by an entire region of points $y$ between $x$ and $z$. The boundary of this region is made by potential solutions to the TPP. It follows that \emph{superadditivity is defined in Euclidean signature by the TPP having a codimension-two manifold of solutions.}

%\hspace{-1cm}
\begin{table}
\begin{center}
{\large Solutions to the TPP: {\bf Euclidean signature}}
\end{center} 
\begin{tabular}{|ccc|c|}
\hline
Character & Example ($\epsilon >0$) & Solutions & Sketch \\ \hline
  \begin{tabular}{c}
  Additive
  \end{tabular} &
  $  d^2(x_1,x_2) =  \deq$  &
  \begin{tabular}{c}
   One point
  \end{tabular} &
  \begin{tabular}{c}
   \includegraphics[width=.8in]{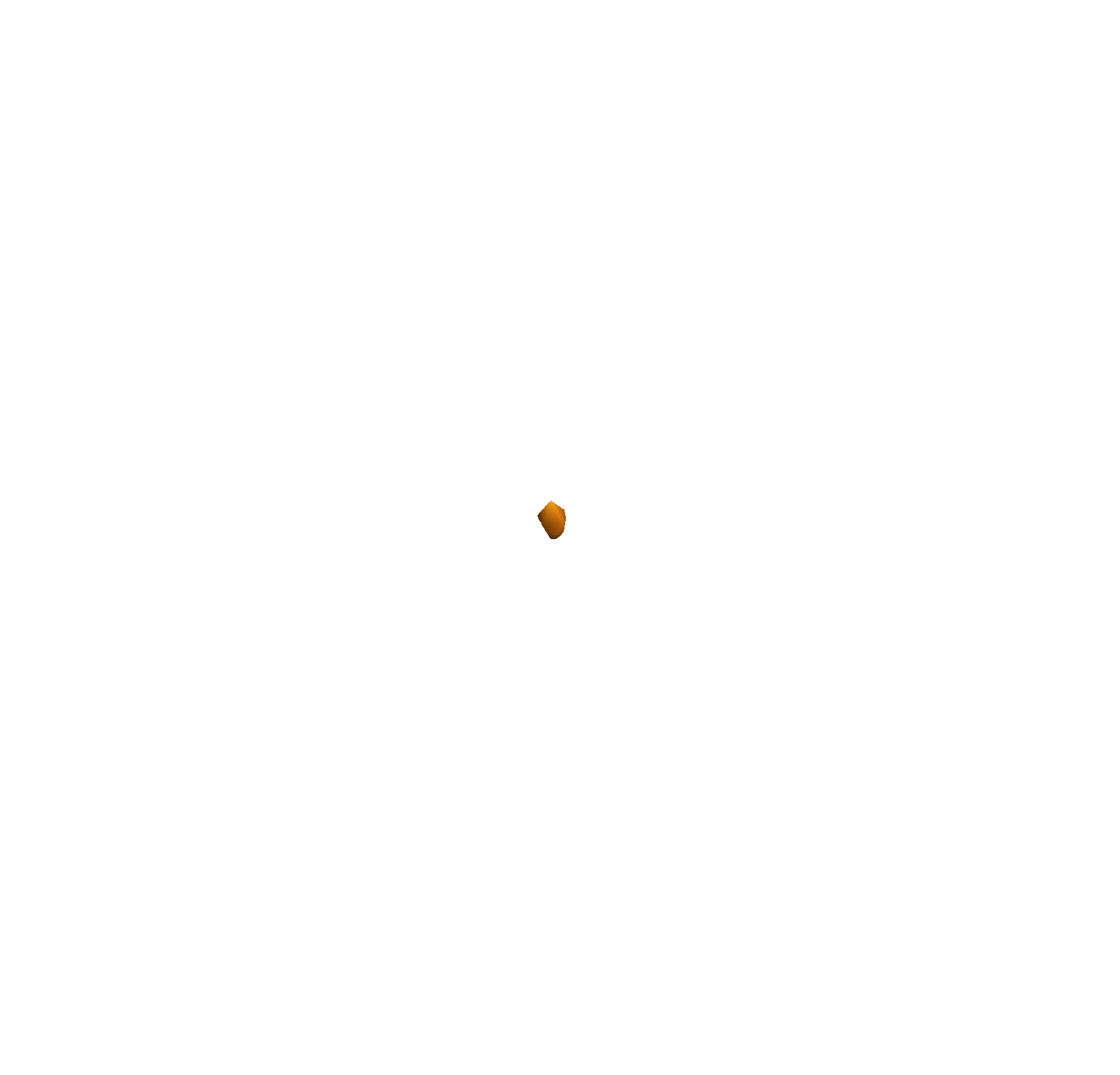} \end{tabular}
  \\
  \hline
  \begin{tabular}{c}
   Subadditive
  \end{tabular} & $d^2(x_1,x_2) = \deq \left[1 - \epsilon \, \deq \right]$  &
  \begin{tabular}{c}
   No solution  \end{tabular} &
  \begin{tabular}{c}
     \includegraphics[width=.8in]{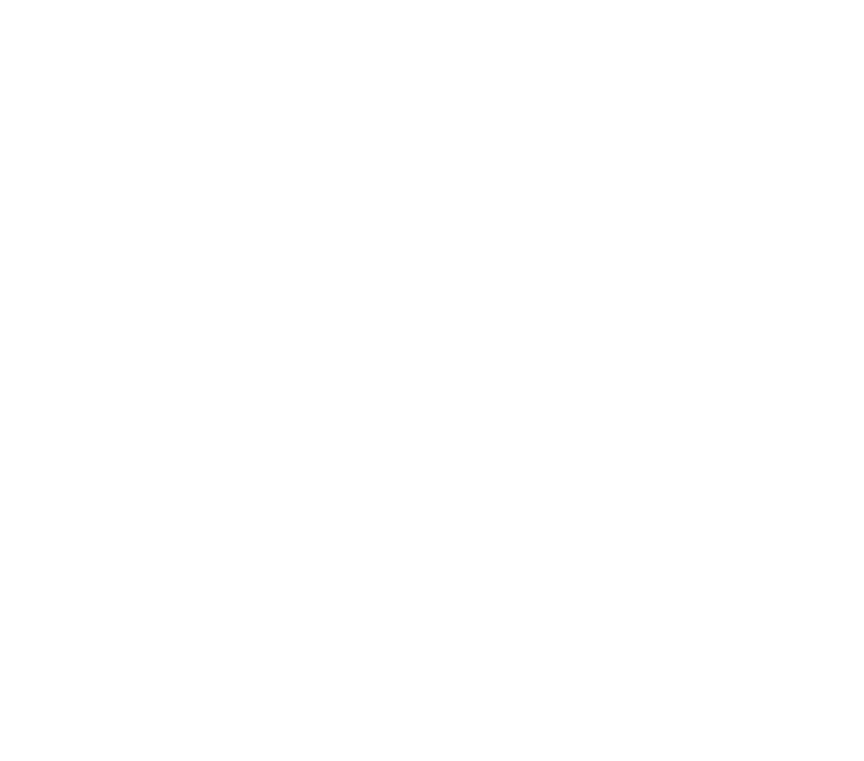}
  \end{tabular} \\
\hline
 \begin{tabular}{c}
   Superadditive
  \end{tabular} & $d^2(x_1,x_2) = \deq \left[1 + \epsilon \, \deq \right]$ &
  \begin{tabular}{c}
    Codimension-two \\ surface    \end{tabular} &
  \begin{tabular}{c}
     \includegraphics[width=.8in]{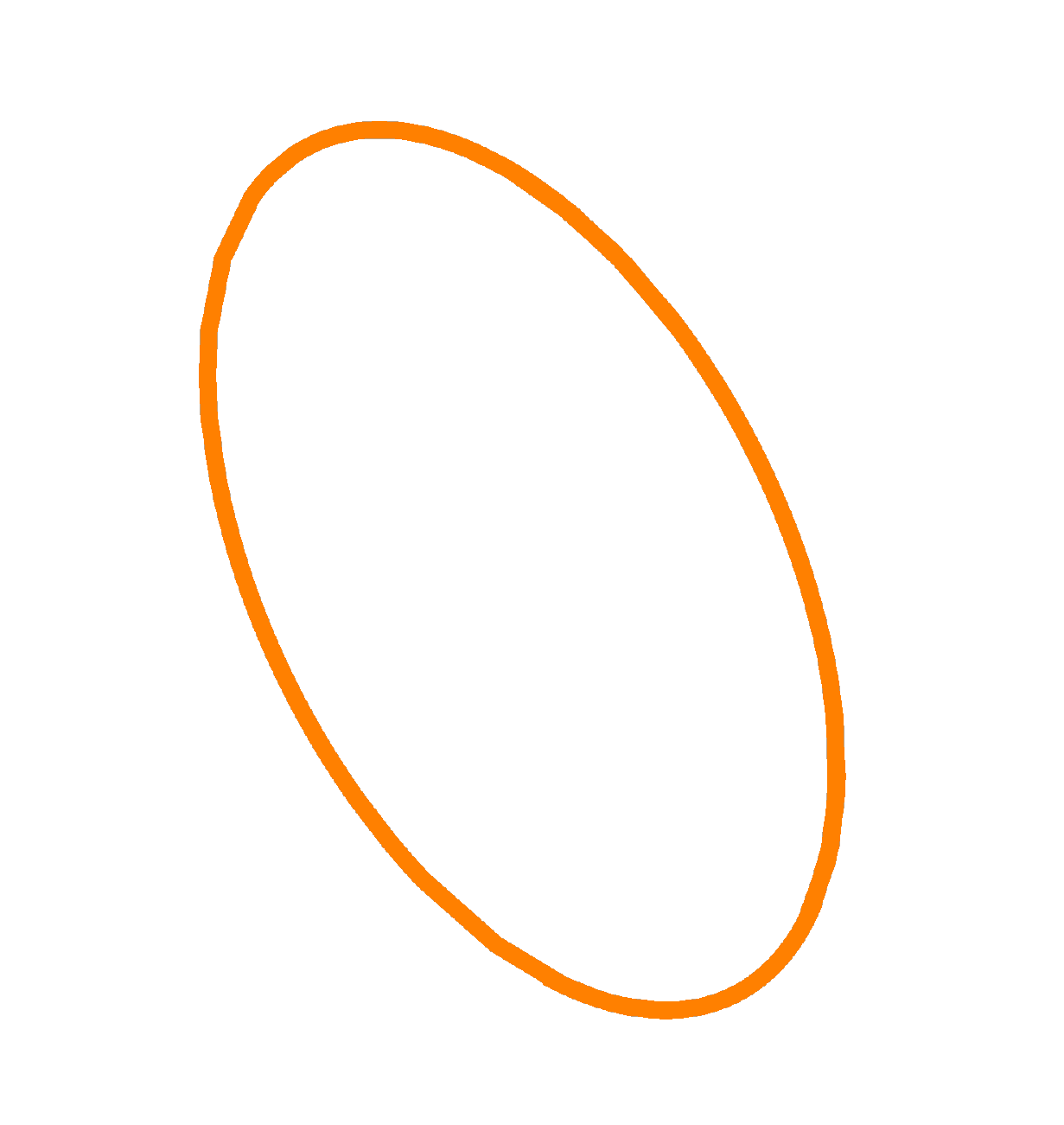}
  \end{tabular} \\
\hline

\end{tabular}

\end{table}

\subsubsection{Local characterization}

By studying the TPP numerically one can easily decide to which of the three above categories some biscalar $d(x,y)$ belongs. A useful local characterization  of $d(x,y)$ can also be made by formulating the TPP while sending $R\rightarrow0$. In this limit $y^i\simeq z^i - \Delta z^i$ and one can expand $d(x,y)$ in condition~\eqref{tpp2} to find
\begin{equation}
R + \frac{\partial}{\partial z^i} d(x,z) \Delta z^i = 0\, .
\end{equation}
Infinitesimal distances in the vicinity of $z$ are governed by the metric tensor in $z$ as defined in~\eqref{metrictensor}. We can thus write condition~\eqref{tpp1} as
\begin{equation}
\Delta z^i = R\,  n^i, 
\end{equation}
with $n^i$ a unit vector in $z$ of unknown direction, $n^i n^j g_{ij}(z) =1$. We obtain that the TPP is stated by the relation 
\begin{equation}
\frac{\partial}{\partial z^i} d(x,z) \,  n^i = -1\, . 
\end{equation}
The above equation can have zero, one, or infinite solutions depending on the modulus of the gradient of $d(x,z)$ in $z$.  If the modulus is less than one the equation has no real solution. If the modulus is one there is precisely one solution, i.e. $n^i$ collinear with $\frac{\partial}{\partial z^i} d(x,z)$ and of opposite  direction. Gradients of modulus larger than one allow a circle of solutions around such a direction. So the character of $d(x,z)$ can be locally determined as follows 
\begin{align} \label{2.11}
\frac{\partial}{\partial z^i} d(x,z) \, g^{ij}(z) \, \frac{\partial}{\partial z^j} d(x,z) = 1\, &:\qquad {\rm additive},\\
\frac{\partial}{\partial z^i} d(x,z) \, g^{ij}(z) \, \frac{\partial}{\partial z^j} d(x,z) < 1\, &:\qquad {\rm subadditive}, \\ \label{2.13}
\frac{\partial}{\partial z^i} d(x,z) \, g^{ij}(z) \, \frac{\partial}{\partial z^j} d(x,z) > 1\, &:\qquad {\rm superadditive}.
\end{align}
One can explicitly verify that the examples~\eqref{ex-euc},~\eqref{ex-2},~\eqref{ex-3} fall in their respective categories.

\subsection{Lorentzian signature}

Lorentzian-signature distances  satisfy the first property $i)$ (symmetry) of metric-spaces, but not the other two.

\subsubsection{Additivity}
We insist to call \emph{additive} the standard geodesic distance associated with a metric tensor. After all, also in Lorentzian signature the geodesic between $x$ and $z$ is made of points $y$  that satisfy $d(x,y)+d(y,z) = d(x,z)$. However, there is no triangle inequality to saturate in Lorentzian signature and the number and type of solutions to the TPP now depend on the nature of $d(x,z)$. As before, we give no strict demonstration of our statements but these can be easily verified by considering the Minkowski distance as a standard example of an additive distance, 
\begin{equation}
d^2(x_1,x_2) = -(\Delta x^0)^2 + \deq \, .
\end{equation}

\emph{If} $d^2(x,z)>0$ the TPP has infinite solutions that span a codimension two ``light cone". Through the vertex of the light cone goes the spatial geodesic between $x$ and $z$. But the TPP conditions can still be satisfied by moving along the locally null directions orthogonal to such a geodesic. The presence of a continuum of points satisfying the TPP is also related with the fact that spacial geodesics are \emph{saddles} of the length functional. 

 \emph{If} $d^2(x,z)<0$ the TPP has strictly one solution, on the timelike geodesic connecting $x$ and $z$. This is related with the fact that timelike geodesics \emph{maximize} the length. Any other point would make the sum of distances smaller, which is the origin of the twin paradox. 

  \emph{If} $d^2(x,z)=0$ the parameter $R$ in the TPP is forced to be zero. The solutions are simply all the points on the null geodesic connecting $x$ and $y$. So in this case the solutions make a one dimensional manifold, as advertised in the introduction. 
 
The infinitesimal version of the TPP is slightly modified in Lorentzian signature but manages to reproduce the same patterns on the tangent space at $z$. 
 
If $x$ and $z$ are at spacelike separation then $\nabla d$ is spacelike and we obtain, as before, the condition\footnote{From now we will often use an obvious simplified notation in which $d(x,z)$ is simply indicated as $d$ and $\partial_\mu$ or $\nabla$ stand for derivatives with respect to $z$.} $n^\mu \nabla_\mu  d   = -1$. It is easy to verify that if $|\nabla d=1|$ such a condition defines a 2+1 dimensional light cone orthogonal to $\nabla d$  on the tangent space.\footnote{On the tangent space one can chose a reference where $\nabla d = (0,1,0,0)$.  $n^\mu \nabla_\mu  d   = -1$ simply implies $n_1 = -1$. The normalization condition for $n$,
\begin{equation}
-n_0^2 + 1 + n_2^2 + n_3^2 = 1
\end{equation}
then defines a two dimensional light cone on the $n_1 =-1$ plane.}
If $x$ and $z$ are timelike $\nabla d$ is an imaginary timelike vector and TPP consists of equations~\eqref{tpp3} and~\eqref{tpp4}. We can write $d = i \tilde d$ so that~\eqref{tpp4} becomes the condition $n^\mu \nabla_\mu  \tilde d   = -1$, with $n$ a unit timelike vector. If $g^{\mu \nu} \nabla_\mu  \tilde d   \nabla_\nu  \tilde d   = -1$ this condition has one and only one solution.\footnote{Chose the reference system such that $\nabla \tilde d = (1,0,0,0)$. The condition $n^\mu \nabla_\mu  \tilde d   = -1$ for a unit timelike vector $n$ has clearly only one solution: $n = (-1,0,0,0)$.}
The null case is degenerate and difficult to reproduce on the tangent space.

\begin{table}
\centering
    \addtolength{\leftskip} {-2cm}
    \addtolength{\rightskip}{-2cm}
\begin{center}
{\large Solutions to the TPP: {\bf Lorentzian signature}}
\end{center} 
\begin{tabular}{|c|cc|c|}
\hline
Character & First two points & Solutions & Sketch \\ \hline
 \begin{tabular}{c}
  \end{tabular} &
 \begin{tabular}{c}
   $d^2(x,z)>0$
  \end{tabular} &
  \begin{tabular}{c}
  Codimension-two \\ surface 
  \end{tabular} &
 \begin{tabular}{c}
   \includegraphics[width=.8in]{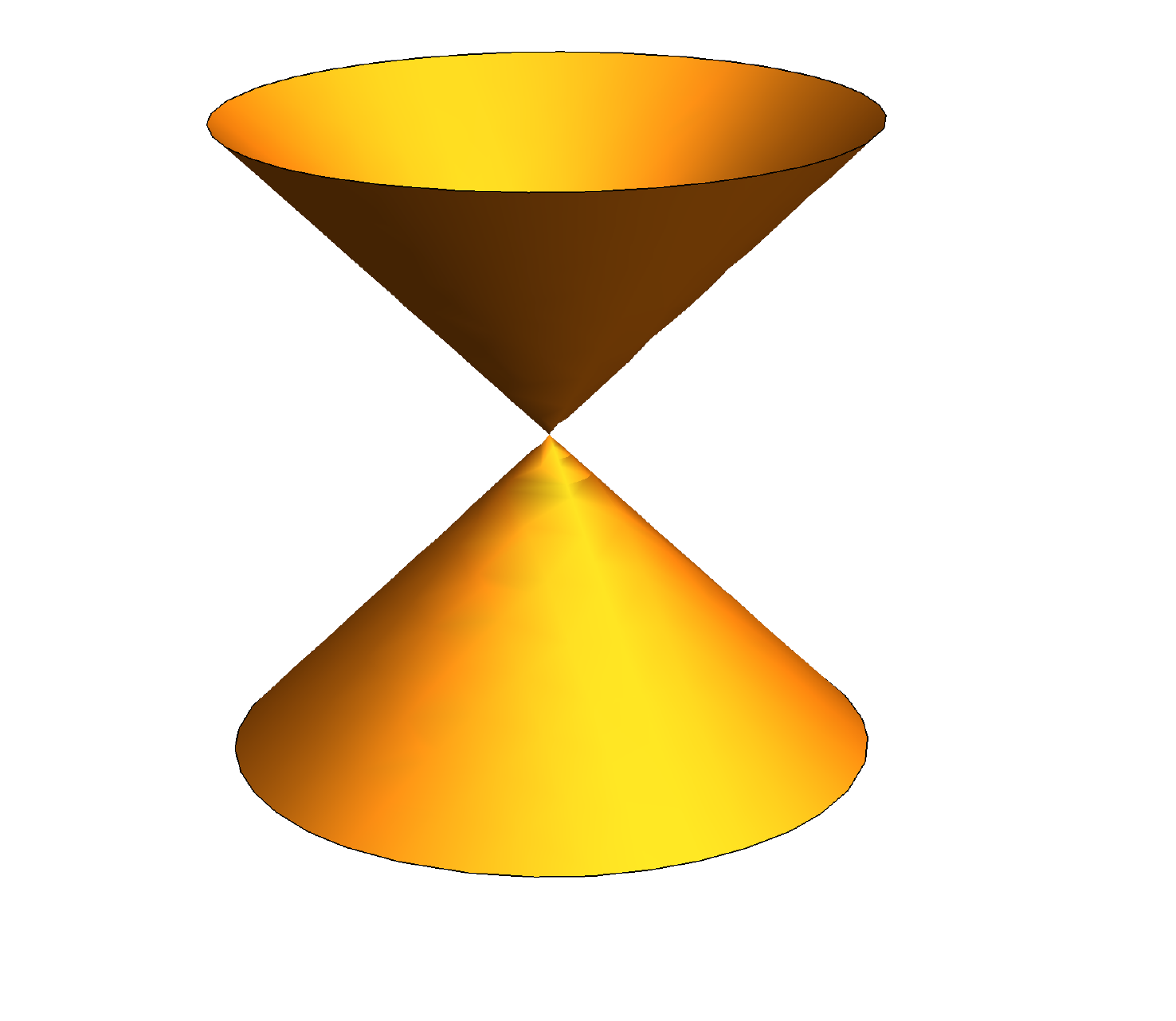} \end{tabular}\\
  \begin{tabular}{c} 
  Additive $(C = 0)$ \\[2mm]
  \emph{e.g.}\ \  \ \ $ d^2(x_1,x_2) = -(\Delta x^0)^2 + \deq$
  \end{tabular} &
    $d^2(x,z)=0$  &
  \begin{tabular}{c}
   One-dimensional \\ curve
  \end{tabular} &
  \begin{tabular}{c}
   \includegraphics[width=.8in]{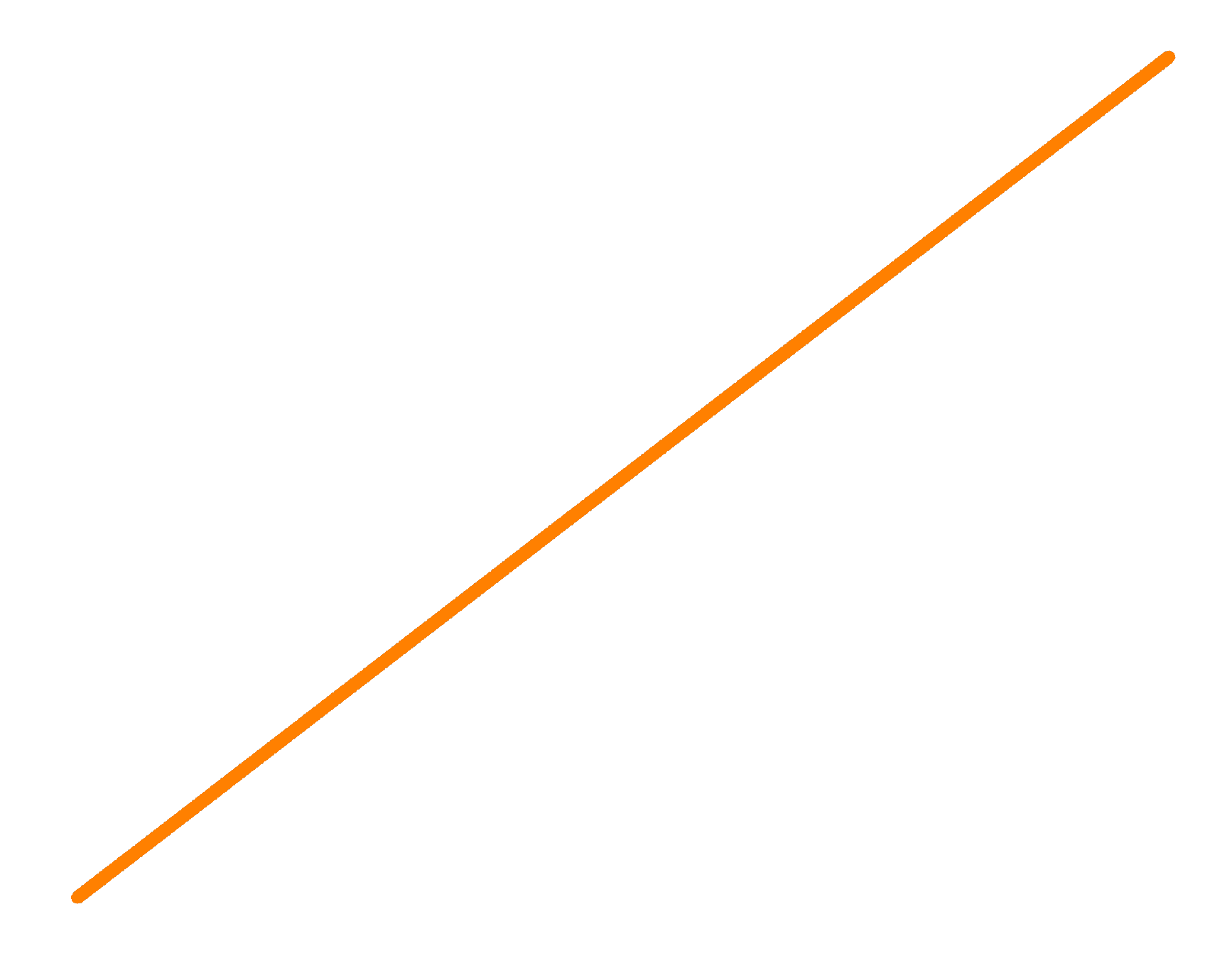} \end{tabular}\\
    \begin{tabular}{c}
  \end{tabular} &
 \begin{tabular}{c}
   $d^2(x,z)<0$
  \end{tabular} &
  \begin{tabular}{c}
 One point
  \end{tabular} &
 \begin{tabular}{c}
   \includegraphics[width=.8in]{onesolu} \end{tabular}
  \\
  \hline
   \begin{tabular}{c}
  \end{tabular} &
 \begin{tabular}{c}
   $d^2(x,z)>0$
  \end{tabular} &
  \begin{tabular}{c}
  Codimension-two \\ surface 
  \end{tabular} &
 \begin{tabular}{c}
   \includegraphics[width=.8in]{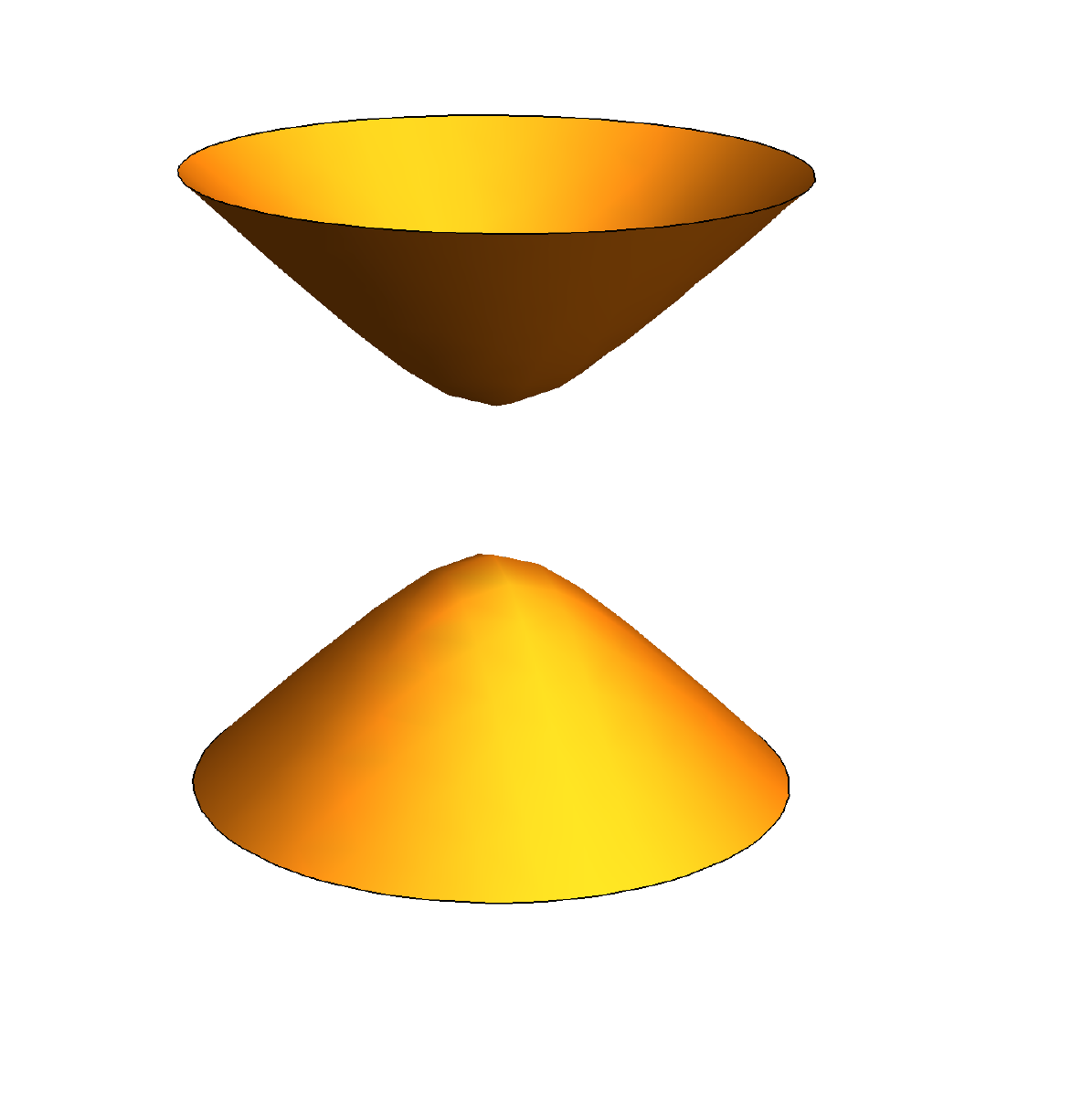} \end{tabular}\\
  \begin{tabular}{c} 
  Subadditive $(C<0)$\\[2mm]
  $ d^2(x_1,x_2) = - (\Delta x^0)^2 + \deq \left[1 - \epsilon \, \deq \right]$
  \end{tabular} &
    $d^2(x,z)=0$  &
  \begin{tabular}{c}
   No solution
  \end{tabular} &
  \begin{tabular}{c}
   \includegraphics[width=.8in]{nosolu} \end{tabular}\\
    \begin{tabular}{c}
  \end{tabular} &
 \begin{tabular}{c}
   $d^2(x,z)<0$
  \end{tabular} &
  \begin{tabular}{c}
 No solution
  \end{tabular} &
 \begin{tabular}{c}
   \includegraphics[width=.8in]{nosolu} \end{tabular}
  \\ \hline 
     \begin{tabular}{c}
  \end{tabular} &
 \begin{tabular}{c}
   $d^2(x,z)>0$
  \end{tabular} &
  \begin{tabular}{c}
  Codimension-two \\ surface 
  \end{tabular} &
 \begin{tabular}{c} \\[-2mm]
   \includegraphics[width=.8in]{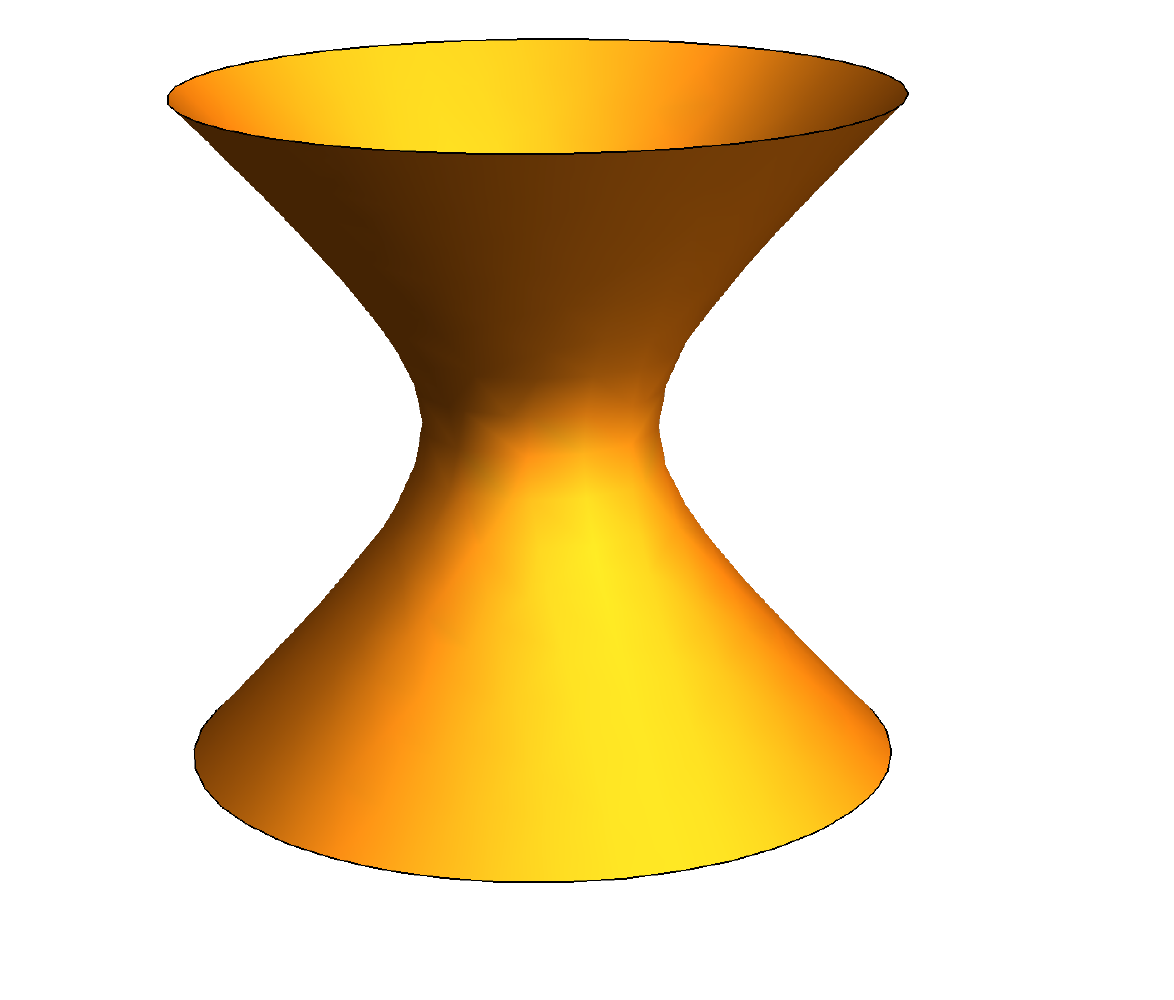} \end{tabular}\\
  \begin{tabular}{c} 
  Superadditive $(C>0)$ \\[2mm]
  $ d^2(x_1,x_2) = - (\Delta x^0)^2 + \deq \left[1 + \epsilon \, \deq \right]$
  \end{tabular} &
    $d^2(x,z)=0$  &
  \begin{tabular}{c}
Codimension-two \\ surface 
  \end{tabular} &
  \begin{tabular}{c}
   \includegraphics[width=.8in]{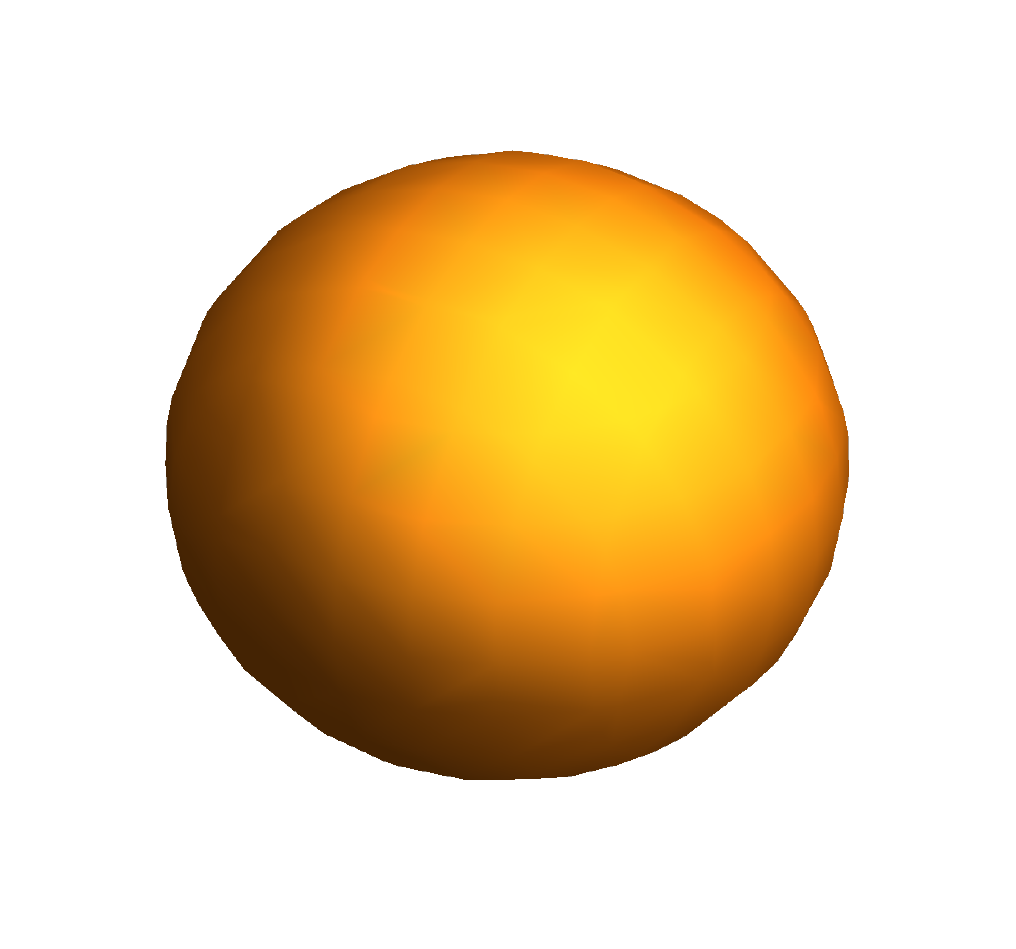} \end{tabular}\\
    \begin{tabular}{c}
  \end{tabular} &
 \begin{tabular}{c}
   $d^2(x,z)<0$
  \end{tabular} &
  \begin{tabular}{c}
Codimension-two \\ surface 
  \end{tabular} &
 \begin{tabular}{c}
   \includegraphics[width=.8in]{sphere} \end{tabular}
  \\ \hline
 \end{tabular} \label{table-lor}
\end{table}

\subsubsection{Spacetime Additivity}

It is clearly more messy having to deal with distances that can turn imaginary, so it is convenient to square them whenever possible. 
The Euclidean conditions~\eqref{2.11}-\eqref{2.13} can be multiplied on both sides by $d^2(x,z)$ to obtain conditions that only involve $d^2(x,z)$. This suggests the introduction of the crucial quantity
\begin{equation} \label{crucial-first}
C(x,y) \equiv \frac14\,  \frac{\partial \,  d^2(x,y)}{\partial y^\mu} \frac{\partial \,  d^2(x,y)}{\partial y^\nu} \  g^{\mu \nu}(y) -  d^2(x,y)\, ,
\end{equation}
where the metric is defined in terms of the distance $d(x,y)$ by \eqref{metrictensor}. We shall dub this the {\it spacetime additivity}. Notice that $C(x,y)$ is not symmetric with respect to $x$ and $y$. For some initial and possibly far away point $x$, $C$ is a real function of $y$, telling how much additivity is violated when playing the TPP with one point fixed at $x$ and the two other points in the vicinity of $y$. 

 In euclidean signature the sign of $C$ directly gives the character of $d$, with additivity simply corresponding to $C=0$ in both signatures. 
So we use $C$ to continue our definition of subadditivity into Lorentzian signature and just define subadditivity as $C<0$ in any signature. A simple example of a subadditive distance in Lorentzian signature is
\begin{equation} \label{ex-4}
d^2(x_1,x_2) = - (\Delta x^0)^2 + \deq \left[1 - \epsilon \, \deq \right]\, ,  
\end{equation}
while superadditivity ($C>0$) is exemplified by 
\begin{equation} \label{ex-4}
d^2(x_1,x_2) = - (\Delta x^0)^2 + \deq \left[1 + \epsilon \, \deq \right]\, . 
\end{equation}

The solutions to the TPP in Lorentzian signature are diverse and summarized in the Table above.

\section{Evaluating C} \label{sec_normal}

In this section we want to express the standard geodesic distance of a metric manifold in a coordinate expansion around a point, and then estimate the additivity $C$ in the presence of a statistical ensemble of metrics. 

\subsection{Normal neighbourhoods}

Let us chose one of the two extremes of $d(x,y)$ as the origin of our coordinate system. From the origin we can send geodesics in every direction to reach any neighboring point and thus calculate the distance $d(0,x)$. Riemann normal coordinates (RNC) $Y^a$ can be of help in this construction because all geodesics passing through the origin can be written as $Y^a = V^a \lambda$, where $V^a$ is a vector tangent at $x=0$ and $\lambda$ is the affine parameter of the geodesic. It follows that in terms of $Y^a$ the distance is simply given as 
\begin{equation}\label{dsquare1}
d^2(0,x) = Y^a \eta_{a b} Y^b\, ,
\end{equation}
where we leave the signature of the flat metric $\eta$ undefined for the moment. 
 
In any coordinate system $x$ one can express the RNC $Y^a$ in terms of the metric and its derivatives calculated at the origin, \footnote{I have replaced `in the origin' with `at the origin' everwhere} 
\begin{equation} \label{ytox}
Y^a = e_\mu^a\left( x^\mu  + \frac12 G^\mu_{ \nu \rho }x^\nu x^\rho + \frac16 G^\mu_{\nu \rho \sigma} x^\nu x^\rho x^\sigma +\dots\right),
\end{equation} 
where
\begin{equation}
 e_\mu^a \eta_{ab} e_\nu^b  = g_{\mu \nu}\, ,\qquad
G^\mu_{\nu \rho} = \Gamma^\mu_{\nu \rho}\, , \qquad
G^\mu_{ \nu \rho \sigma} = \partial_{(\nu} \Gamma^\mu_{\rho \sigma)} + \Gamma^\mu_{\alpha (\nu}\Gamma^\alpha_{\rho \sigma)}\, .
\end{equation}
All quantities on the RHS are evaluated at the origin and in the coordinates $x$.  The above expansion can be inserted into~\eqref{dsquare1} to obtain the distance in any coordinate system, 
\begin{align} \label{dsquare2}
d^2(0,x) =& \ g_{\mu \nu}x^\mu x^\nu + \frac12 g_{\mu \nu, \rho} \, x^\mu x^\nu x^\rho 
 - \frac{1}{12}\left(g_{\alpha \beta}\Gamma^\alpha_{\mu \nu} \Gamma^\beta_{\rho \sigma} - 2 g_{\mu \nu, \rho \sigma}\right) x^\mu x^\nu x^\rho x^\sigma + {\cal O}(x^5)\, .
\end{align}
It is a lengthy but straightforward calculation to check that the spacetime additivity $C$ defined in~\eqref{crucial-first} identically vanishes for the above distance. This is expected because~\eqref{dsquare2} is a geodesic distance. In quantum gravity, for a certain coordinate system of physical observers $x$, the above expression and each coefficient of the $x$-expansion becomes an operator. We can thus generally express ${\bar d}^{\, 2}(0,x) \equiv \langle d^2(0,x) \rangle$ by substituting the coefficients in~\eqref{dsquare2} with expectation values, $g_{\mu \nu}\rightarrow \langle g_{\mu \nu}\rangle \equiv \bar g_{\mu \nu}$ etc.

The spacetime additivity calculated for $\bar d$ then reads
\begin{equation} \label{crucial-average}
C(0,x) = - \frac14\left( \langle g_{\alpha \beta}\Gamma^\alpha_{\mu \nu} \Gamma^\beta_{\rho \sigma}\rangle - \bar g^{\alpha \beta}\langle \Gamma_{\alpha \mu \nu} \rangle \langle \Gamma_{\beta \rho \sigma} \rangle \right) x^\mu x^\nu x^\rho x^\sigma + {\cal O}(x^5)\, ,
\end{equation}
where $\Gamma_{\alpha \mu \nu} \equiv g_{\alpha \beta} \Gamma^\beta_{ \mu \nu}$ and we remind the reader that $\bar g^{\mu \nu}$ is the inverse of $\langle g_{\mu \nu}\rangle = \bar g_{\mu \nu}$. All operators are evaluated at the origin. 
Notice that $C$ starts at order ${\cal O}(x^4)$ and, as announced, it vanishes when averages are neglected, i.e., on a classical spacetime. 

It is possible to write the same quantity as a ``square", 
\begin{equation}
C(0,x) = -\frac14 \langle Q_a \eta^{ab} Q_b\rangle\, ,
\end{equation}
where
\begin{equation}
Q_a = \left(e^\alpha_a \, \Gamma_{\alpha \mu \nu} - e_{\beta a} \, \bar g^{\alpha \beta} \langle \Gamma_{\alpha \mu \nu}\rangle \right)x^\mu x^\nu\, ,
\end{equation}
where $e^\alpha_a$ are tetrad \emph{operators} at the origin, $e^\mu_a \eta^{ab} e^\nu_b = g^{\mu \nu}$ etc. 

\subsection{Average Euclidean distances are always subadditive} \label{subsec-euc}

In Euclidean signature $\eta^{ab} \rightarrow \delta^{ab}$, so from the above it is clear that $C\leq 0$: $\bar d(x,y)$ is subadditive. In other words, the root mean square of Euclidean geodesic distances is a biscalar $\bar d(x,y)$ which is still a distance in the sense of metric-spaces, i.e. it satisfies the triangle inequality. However, it will generally fail to saturate it. 
This is one of three pieces of evidence that we give of this. One limitation of the above proof is that $C$ might turn positive at order higher than ${\cal O}(x^4)$. In App.~\ref{app-resum} we show that $C\leq 0$ to any order in the coordinate expansion. However, we are able to do so only 
for a classical statistical mixture of two geodesic distances. In App.~\ref{app-triangle} we simply prove that the average of a classical statistical mixture of any number of distances satisfying the triangle inequality also satisfies the triangle inequality. Indirectly, this is a proof that such an average is subadditive, as superadditive distances just fail this inequality. One could argue that in Euclidean signature it is not a restriction to consider classical statistics, because the three-dimensional metric $h_{ij}$ commutes with all its spatial derivatives and thus the distance operator is diagonal on the eigenbasis of the distance (see also~\cite{Piazza:2021ojr} on this). 

\subsection{Lorentzian signature} \label{subec-lorentzian}

In Lorentzian signature there is no proof of subadditivity. Quite the opposite, there exists explicit examples of superadditivity (see Sec.~\ref{FRW} below).
Situations where $C(x,y)$ is both negative and positive depending on the locations of the two extremes are also possible in principle. In this respect, since we are mostly concerned with causality, it is interesting to evaluate $C(x,y)$ around the region where $\bar d(x,y) = 0$.

In the Lorentzian case we are in particular ultimately interested in quantum expectation values, be they in a pure state or a mixed state. Since the metric itself is quantized, the distance $\bar d(x,y) $ is itself a non-linear and non-trivial function of the metric. As such at the quantum level it is a composite operator and its very definition is subtle. In perturbation theory, the leading divergent contribution to $d(x,y)$ comes from vacuum fluctuations in Minkowski, and this term may be removed by normal ordering. It is the beyond the scope of our present discussion to consider the likely higher order composite operator renormalizations.

As discussed in Sec.~\ref{sec_observers}, for hypersurface-orthogonal geodesic observers the metric in unitary gauge takes the form
\begin{equation}
g_{\mu \nu} dx^\mu dx^\nu = -dt^2 + \gamma_{ij} dx^i dx^j\, 
\end{equation}
so that the averages hit the spatial metric $\gamma_{ij}$ and its time derivatives. From~\eqref{crucial-average} we obtain
\begin{align} \label{3.11}
C  = & \ \frac14\left [ \frac14\left(\langle \dot \gamma_{ij} \dot \gamma_{lk} \rangle - \langle \dot \gamma_{ij} \rangle\langle \dot \gamma_{lk} \rangle  \right) x^i x^j x^k x^l \right. \\ \label{3.12}
& - \left(\langle \gamma^{pq} \dot \gamma_{p i} \dot \gamma_{q j}\rangle - \bar \gamma^{pq} \langle \dot \gamma_{p i} \rangle \langle \dot \gamma_{q j}\rangle\right)  t^2 x^i x^j \\  \label{3.13}
&- 2 \left( \langle \Gamma^p_{ij} \dot \gamma_{p k}\rangle - \bar \gamma^{pq} \langle \Gamma_{p i j} \rangle \langle \dot \gamma_{q k}\rangle\right)  t \, x^i x^j x^k\\
&- \left. \left(\langle \gamma^{pq} \Gamma_{p ij} \Gamma_{q kl}\rangle - \bar \gamma^{pq}  \langle \Gamma_{p ij}  \rangle \langle \Gamma_{q kl}\rangle \right) x^i x^j x^k x^l \right]\, , \label{3.14}
\end{align}
where, again, $\bar \gamma^{ij}$ is the inverse of $\bra \gamma_{ij}\ket$. 
One thing to notice is that average distances along the $t$ direction are strictly additive. In other words, $C=0$ if $x^i=0$.  This is because we are using the proper times of the observers at $\vec x$ = const. to \emph{define} time. The proper time of our observers---or, the direction probed by their free-falling---is additive by construction.

The last line~\eqref{3.14} is negative because it is just the expression for $C$ in Euclidean signature. The second line can be proven to be negative in a way similar to the Euclidean case already discussed, 
\begin{equation} \label{reason1}
-\left(\langle \gamma^{pq} \dot \gamma_{p i} \dot \gamma_{q j}\rangle - \bar \gamma^{pq}  \langle \dot \gamma_{p i} \rangle \langle \dot \gamma_{q j}\rangle\right)  =  -\, \bra Q^p Q_p\ket\, ,
\end{equation}
with
\begin{equation} \label{reason2}
Q_p \equiv \left(e^j_p \, \dot \gamma_{j i} - e_{j p}\, \bar \gamma^{j k} \bra \dot \gamma_{k i}\ket\right) x^i\, ,
\end{equation}
and $\gamma^{pq}e^i_p e^j_q = \delta^{ij}$. 
The third line~\eqref{3.13} gives a contribution to $C$ whose sign is (space- and time-) parity odd. In homogeneous or isotropic situations we expect this term to vanish. 
The first line's contribution, however, is always positive. One can try to see what happens around the region where $\bar d(0,x) =0$, which is the most interesting for causality. To this order it is sufficient to substitute $t^2= \bra \gamma_{ij}\ket x^ix^j$ in the second and third lines. 

Apart from \emph{ad hoc} situations we find that the negative contributions to $C$ tend to overwhelm the positive one. 
In the following section we provide several examples when this is the case and one counterexample.

\section{Examples}

With the formula~\eqref{3.11}-\eqref{3.14} at our disposal we can calculate $C$ in different situations. 

\subsection{Plane Waves} \label{sec_planewaves}

Let us go back to the very first example of the introduction and consider Minkowski space traversed at by a superposition of planar gravitational waves of different polarizations. The formulas of last section are easily applied to this case. Let us consider a metric of the type 
\be \label{null-metric}
ds^2 =- dt^2 + dz^2 + H_{MN}(t-z,x^M) dx^M dx^N \, ,
\ee
where $M$, $N = 1,2$ are the two transverse coordinates. This \emph{is not} the most general spacetime with a null Killing vector field but it is general enough to contain all plane waves with wavefront at $t = z$.  In this case $H_{MN}$ does not depend on $x^M$ and this can be a vacuum solution of the Einstein equations or it can be supported by a massless field. All these cases are considered in some detail in App.~\ref{grav-waves}. 

By specializing~\eqref{3.11}-\eqref{3.14} to the metric~\eqref{null-metric} one obtains 
\begin{align}
C  = & -\frac14\left [  \left(\langle H^{LK} H'_{L M}  H'_{K N}\rangle - \bar H^{LK} \langle  H'_{LM} \rangle \langle H'_{KN}\rangle\right) (t^2 - t z+z^2) x^M x^N \right. \\
&+  \left(\langle H^{PQ} \Gamma_{P MN} \Gamma_{QKL}\rangle - \bar H^{PQ}  \langle \Gamma_{PMN}  \rangle \langle \Gamma_{QKL}\rangle \right) x^M x^N x^K x^L \\
&+ \left. 2 \left(\bra \Gamma^P_{I  J} H'_{PK}\ket - \bar H^{PQ} \bra \Gamma_{PIJ}\ket \bra H'_{QK}\ket\right) x^I x^J x^K (t-z)
\right]\, .
\end{align}
Apart from the parity-odd term in the last line, the other two contributions to $C$ are negative. The first line, because 
\begin{equation} 
 t^2 - t z+z^2 = \left(t - \frac{ z}{2}\right)^2 + \frac34 z^2\, , 
 \end{equation}
 and for the same reasons explained at eqs.~\eqref{reason1}-\eqref{reason2}. The second line, again, is a purely Euclidean contribution so it also contributes negatively  to $C$. In all situations where the third line is null for symmetry reasons $C$ is negative definite. More importantly, for plane waves, $H_{MN}$ does not depend on $x^M$, all transverse Christoffel symbols $ \Gamma^P_{I  J}$ vanish, together with the second and third lines of the above expression. So for a superposition of plane waves sharing the same wavefront we always have $C\leq0$.

Despite the concise formulas for $C$, we also find instructive to solve the full \emph{third point problem} ``hands on" with the aid of Mathematica. This is how the problem has been tackled at the beginning of this research. We give account of this numerical analysis in App.~\ref{grav-waves}, where also plane waves in scalar-gravity are considered. In all these cases we find that average distances are \emph{subadditive}.

\subsection{FRW geometries} \label{FRW}

It is straightforward to calculate $C$ in the case where the metric is taken to be a homogenous and isotropic spatially flat geometry\begin{equation}
C = -\frac14 \left(4 \langle a^2  \rangle \langle \dot a^2 \rangle - 3 \langle a \dot a \rangle^2 - \langle a^2 \dot a^2 \rangle \right ) x^4\, 
\end{equation}
at $d(0,x) = 0$. It will prove useful to write this as
\begin{equation}
C = -\frac14 \left(3 \left[ \langle a^2  \rangle \langle \dot a^2 \rangle -  \langle a \dot a \rangle^2\right] + \left[\langle a^2  \rangle \langle \dot a^2 \rangle- \langle a^2 \dot a^2 \rangle \right] \right)x^4\, .
\end{equation}
The first expression can easily seen to be positive by minimizing the positive quantity $\langle (\dot a - \lambda a)^2 \rangle \ge 0$ over $\lambda$. Without any other input, it is not possible to argue that $C$ is negative semi-definite since the second term in brackets can take any sign. Crucially, the dynamics of the system relates $\dot a$ with $a$ via the Friedman equation. This is true not just classically, but at the quantum level via the Heisenberg equations of motion. Let us consider the case in which the matter is taken to be a fluid of fixed equations of state $w=p/\rho$. The Friedman equation then takes the form
\be
\dot a^2 = A \, a^{-\gamma} \, ,
\ee
where  $\gamma = 1+3 w$.
Consider now the combination
\be
\langle a^2  \rangle \langle \dot a^2 \rangle- \langle a^2 \dot a^2 \rangle  = \langle a^2  \rangle \langle a^{-\gamma} \rangle- \langle a^{2-\gamma}\rangle
\ee
The RHS can be shown to be positive by using the following trick. 
Let us define $\phi(t) \equiv \bra a^t\ket^{1/t}$ for any operator $a$. By generalizing the proof given in~\cite{norris} it is easy to show that $\phi(t)$ is a monotonically increasing function.\footnote{Define $F(t)\equiv t^2 \phi'(t)/\phi(t)$. We have 
\be 
F'(t) = \frac{t}{\bra a^t\ket^2} \left(\bra a^t\ket \bra a^t(\ln a)^2 \ket - \bra a^t \ln a \ket^2\right)\, .
\ee
By the Cauchy-Schwartz inequality the term in parenthesis is positive, so $F'(t)$ has the same sign of $t$. This means that $F(t)$ has a minimum at $F(0) = 0$. So $F(t)$ and $\phi'(t)$ are positive for any $t\neq0$. 
}

We can write  
\begin{equation}
\langle a^2  \rangle \langle a^{-\gamma} \rangle- \langle a^{2-\gamma}\rangle = \phi(2)^2 \phi(- \gamma)^{-\gamma} - \phi(2 -  \gamma)^{2-\gamma}\, .
\end{equation}
Now, if $\gamma >0$, $\phi(2)>\phi(2-\gamma)$. Which means 
\be
 \phi(2)^2 \phi(- \gamma)^{-\gamma} - \phi(2 -  \gamma)^{2-\gamma} >  \phi(2-\gamma)^2 [\phi(- \gamma)^{-\gamma} - \phi(2 -  \gamma)^{-\gamma} ] >0\, ,
 \ee
the last inequality coming from the fact that inside the square bracket $\phi(- \gamma) < \phi(2 -  \gamma)$, which implies $\phi(- \gamma)^{-\gamma} > \phi(2 -  \gamma)^{-\gamma}$ for $\gamma>0$. 
This implies
\be
 \langle a^2  \rangle \langle a^{-(1+3 w)} \rangle- \langle a^{(1-3 w)}  \rangle  \ge 0 \, , \quad  w\ge -\frac{1}{3} \, .
\ee
hence
\be
C\le 0 \, , \quad  w\ge -\frac{1}{3} \, .
\ee
It is easy to see that this proof cannot be extended to the case of accelerating geometries $w\le -1/3$. For instance, in the special case $w=-1$ in which we are averaging over de Sitter geometries then
\be
C = - A^2 \left( \langle a^2 \rangle^2- \langle a^4 \rangle \right) \ge 0 \, . 
\ee
\\
The fact that decelerating FRW geometries appear to lead to $C\le 0$ as in the case of Euclidean geometries is paralleled by another interesting observation. The naive Wick rotation of an FRW geometry only leads to an allowable complex metric in the sense of Kontsevich and Segal \cite{Kontsevich:2021dmb} and Louko and Sorkin \cite{Louko:1995jw} for a decelerating geometry. Taking the explicit form of the FRW geometry in proper time
\be
d s^2 =- dt^2 + t^{\frac{2}{3(1+w)}} d \vec x^2 \, ,
\ee
under a Wick rotation $t= e^{-i \pi/2} \tau$ this leads in general to a complex geometry
\be
d s^2 =d \tau ^2 + e^{-i \pi/(3(1+w))}\tau^{\frac{2}{3(1+w)}} d \vec x^2 \, .
\ee
The condition that the complex metric is allowable, which amounts to the requirement that the path integral for scalar and $p-$form matter converges under the Wick rotation, is that \cite{Witten:2021nzp}
\be
\sum_{i=1}^4 |Arg(\lambda_i)|< \pi \, ,
\ee
where $\lambda_i$ are the eigenvalues of the metric. In the present case this is
\be
3 \times \frac{2}{3(1+w)}\pi  < \pi \,
\ee
which is equivalent to the requirement that $w \ge -1/3$.

\subsection{Gravitational perturbations on homogeneous background}

Let us expand the metric around a classical (possibly time dependent) homogeneous value,
\begin{equation}
\gamma_{ij} = a^2(t) \left(\delta_{ij} + h_{ij} \right)\, . 
\end{equation}
By assumption the one point functions of $h_{ij}$ vanish, and we have
\begin{align} 
C  = & \ \frac14\left[ \frac14 \langle \dot {h}_{ij} \dot {h}_{kl} \rangle x^i x^j x^k x^l - \langle \dot {h}_{p i} \dot {h}_{p j}\rangle  \Delta t^2 x^i x^j \right. \\   
& + 2 H \,  \langle {h}_{p i} \dot {h}_{p j}\rangle  \Delta t^2 x^i x^j \\
&- \left(2 \langle \dot {h}_{pk} {h}_{i p,j}\rangle -\langle \dot {h}_{pk} {h}_{i j,p}\rangle \right) \Delta t\,  x^i x^j x^k\\ 
&-  \left. \left(\langle {h}_{i p,j}{h}_{k p,l} \rangle - \langle {h}_{i p,j}{h}_{k l,p} \rangle +\frac14 \langle {h}_{i j,p }{h}_{k l,p} \rangle \right) x^i x^j x^k x^l \right]\, , 
\end{align}
where we have expanded around a point $t = t_0$, $\vec x = 0$, set $a(t_0) = 1$ and $H = \dot a/a$. 
The above expression can be evaluated in perturbation theory, with some prescription (e.g. normal ordering) in order to subtract the infinite vacuum contribution. Since we are especially interested in understanding what happens along the light cone we set $x^i = u^i \Delta t$, with $u^i$ a unit spatial vector. This gives 

\begin{align} \label{c1}
C  = & \ \frac{\Delta t^4}{4}\left[ \frac14 \langle \dot {h}_{ij} \dot {h}_{kl} \rangle u^i u^j u^k u^l - \langle \dot {h}_{p i} \dot {h}_{p j}\rangle  u^i u^j \right. \\   \label{c2} 
& + 2 H \, \langle {h}_{p i} \dot {h}_{p j}\rangle   u^i u^j \\ \label{c3}
&- \left(2 \langle \dot {h}_{pk} {h}_{i p,j}\rangle -\langle \dot {h}_{pk} {h}_{i j,p}\rangle \right)  u^i u^j u^k\\  
&- \left. \left(\langle {h}_{i p,j}{h}_{k p,l} \rangle - \langle {h}_{i p,j}{h}_{k l,p} \rangle +\frac14 \langle {h}_{i j,p }{h}_{k l,p} \rangle \right) u^i u^j u^k u^l \right]\, . \label{c5}
\end{align}
With the only possible exception of line~\eqref{c2} all terms above either vanish or are negative, as we now show. 
By homogeneity, line~\eqref{c3} vanishes. Line~\eqref{c5} is purely spatial and thus gives a negative contribution to $C$. 

The first line can be evaluated as follows. We can decompose $h$ into tensors and scalars. In the case of tensors  tracelessness and isotropy  constrain the form of $\langle \dot {h} \dot {h} \rangle$ to
\begin{equation} 
\langle \dot {h}_{ij} \dot {h}_{kl} \rangle = A( 3 \delta_{ik}\delta_{jl} + 3 \delta_{il}\delta_{jk}-2 \delta_{ij}\delta_{kl}),
\end{equation}
for some positive constant $A$. But then we obtain for the first line~\eqref{c1}
\begin{equation}
- \frac14 \langle \dot {h}_{ij} \dot {h}_{kl} \rangle u^i u^j u^k u^l + \langle \dot {h}_{p i} \dot {h}_{p j}\rangle  u^i u^j = 9 A.
\end{equation}
So all contributions to $C$ are negative, except possibly for line~\eqref{c2}. The latter is vanishing on a Minkowski background. Moreover, schematically, $ : h \dot h : \ \sim \ :(\alpha + \alpha^\dagger)(\alpha - \alpha^\dagger):$, with $\alpha$ the annihilator operator, and thus it clearly vanishes also on a thermal state. Even when positive, it is unlikely that~\eqref{c2} can counterbalance all other negative contributions. 

In App.~\ref{app-thermal} these arguments are explicitly checked by calculating $C$ for a thermal state of gravitons in Minkowski space at temperature $T$. We obtain (see the Appendix for the correct numerical factors)
\begin{equation}
C \simeq - \Delta t^4 \frac{T^4}{M^2_{Pl}}\, .
\end{equation}
Indeed, $C$ has dimensions of a length squared. Non additivity becomes sizable at distances of order $ \ell \sim M_{Pl}/T^2$. If the gravitons were in thermal equilibrium during radiation domination, for example, this would be roughly the size of the Hubble horizon at that epoch. 

For gravitons in a cosmological Bunch Davies vacuum, naive dimensional analysis suggests that this effect is severely suppressed. For example, at the end of inflation we expect the relevant length scale to be of order $ \ell \sim M_{Pl}/H_*^2$, with $H_*$ the Hubble parameter. 

A similar qualitative analysis can be done for scalar fluctuations. In this case homogeneity implies 
\begin{equation} 
\langle \dot {h}_{ij} \dot {h}_{kl} \rangle = B \delta_{ij}\delta_{kl},
\end{equation}
for some positive $B$.
Again, we have an overall negative contribution to $C$, as the first line~\eqref{c1} gives
 \begin{equation}
- \frac14 \langle \dot {h}_{ij} \dot {h}_{kl} \rangle u^i u^j u^k u^l + \langle \dot {h}_{p i} \dot {h}_{p j}\rangle  u^i u^j = \frac34 B. 
\end{equation}
In the Bunch Davies vacuum during inflation the effect is of the same order as that of gravitons, with the known enhancement of the inverse  slow-roll parameter $1/\epsilon$.

\section{Causality of average distances}\label{qp}

We have shown that, once a class of free-falling observers and their associated coordinates $x^\mu$ are identified on a quantum ensemble of metrics, the average distance $\bar d(x,y)$  is a well-defined calculable quantity. But what is the meaning of $\bar d(x,y)$? 

\subsection{The meaning of $\bar d(x,y)$} \label{sec_meaning}
We argued in the introduction that $\bar d(x,y)$ should be a fair proxy for the causal relations among the observers.  A rigorous approach to the problem would be to consider the interactions due to some physical field between two observers. For instance, suppose $\phi(x)$ denotes some physical field, e.g. photon, Higgs. The act of creating the associated particle localized near a point $x_1$ in unitary gauge can be associated by the addition of a classical source $J(x)$ which is peaked near $x_1$ via the interaction
\be
S_1 = \int d^4 y \sqrt{-g(y)} \phi(y) J(y) \, .
\ee
Assuming the observer localized at $x$ can measure the in-in expectation value of the field we have for a fixed geometry\footnote{$\bar T$ is the anti-time ordering operator which arises in the in-in (Schwinger-Keldysh/CTP) formalism.}
\be
\langle  \phi(x) \rangle = \langle 0| \bar T e^{-i \int d^4 y \sqrt{-g(y)} \phi(y) J(y)}    T \left[ \phi(x) e^{i \int d^4 y \sqrt{-g(y)} \phi(y) J(y)} \right] |0 \rangle \ee
which in the linear response (weak source) regime is
\be \label{weakregime}
\langle  \phi(x) \rangle = i \int d^4 y \sqrt{-g(y)} \theta(x^0-y^0) \langle 0| [\phi(x),\phi(y)]|0 \rangle  J(y)\, .
\ee

For a fixed geometry, causality tells us that $[\phi(x),\phi(y)] $ vanishes for spacelike separations. By contrast when we further average over the metric either by accounting for metric fluctuations or for an ensemble of semi-classical metrics, then what is relevant in linear response theory is the expectation value of the retarded propagator $G_{\rm ret}(x,y) = -i \theta(x^0-y^0) \sqrt{-g}   \langle 0 | [\phi(x),\phi(y)]|0 \rangle$ 
\be
\bar G_{\rm ret}(x,y) = \langle G_{\rm ret}(x,y) \rangle \, .
\ee
Crucially $\bar G_{\rm ret}(x,y) $ is not the retarded propagator of the average metric nor of any single effective metric and in this sense shares the same failure of additivity. For a statistical ensemble of metrics, $\bar G_{\rm ret}(x,y) $ only vanishes when $x$ and $y$ are spacelike separated with respect to every metric in the ensemble. For a quantum superposition no such simple statement can be made. As is well known, the retarded propagator on a curved spacetime can be approximated as a function of the geodesic distance $d(x,y)$.
In particular for a massless field, the retarded propagator in four dimensions has the Hadamard form \cite{DeWitt:1960fc,Harte:2012uw}
\be\label{retarded}
 G_{\rm ret}(x,y) = \theta(x^0-y^0)  \left[ \frac{1}{2 \pi}\sqrt{\Delta (x,y)} \delta(d^2(x,y))- 2\pi v(x,y) \theta(-d^2(x,y)) \right]
\ee
where the first term has support on the light cone and the second term is the tail inside the past light cone which generically arises in curved spacetimes. Here $\Delta(x,y)$ is the van Vleck-Morette determinant which is itself determined by $d^2(x,y)$  \cite{Visser:1992pz} 
\be
\Delta(x,y) = \frac{\det(\partial^x_{\mu} \partial^y_{\nu} d^2(x,y))}{2 \sqrt{-g(x)} \sqrt{-g(y)}}\, .
\ee
such that $\Delta(x,x)=1$. In this sense, what is relevant for causality is the expectation value of a function of the squared geodesic distance. Unfortunately the distributional nature of this function makes it particularly challenging to perform this computation which is why we have resorted to evaluating the average of $d^2(x,y)$ alone.

It is interesting to compare our discussion here with the proposal of \cite{Saravani:2015moa} (see also \cite{Perche:2021mue}) of using the Feynman or Wightman propagator on curved spacetimes as a means to infer an effective quantum metric. The idea is to use the fact that on a fixed geometry the singular structure of the Wightman or Feynman propagator analogous to \eqref{retarded} is given by the squared geodesic distance 
\be \label{eq_W}
 G_F(x,y) = \sqrt{\Delta (x,y)} \frac{1}{4 \pi^2} \frac{1}{d^2(x,y)+i \epsilon}- v(x,y) \ln(d^2(x,y) +i \epsilon) +w(x,y)
\ee
and this relation can be inverted to give an equation for the metric
\be\label{effmetric}
\tilde g_{\mu\nu}(x)=-\frac{1}{8 \pi^2} \lim_{y \rightarrow x} \frac{\partial}{\partial x^\mu} \frac{\partial}{\partial y^\nu}  G_F(x,y)^{-1} =-\frac{1}{8 \pi^2} \lim_{y \rightarrow x} \frac{\partial}{\partial x^\mu} \frac{\partial}{\partial y^\nu}  W(x,y)^{-1} \, .
\ee
which is analogous to our definition of the average metric tensor \eqref{metricave}. The simplicity of this result stems from the fact that in the coincidence limit it is the leading $1/d^2(x,y)$ singularity which dominates the correlation functions and by the equivalence principle applied in a locally inertial frame guarantees that this is the same as it is in Minkowski spacetime. The covariance of its definition then ensures that it is equivalent to the metric at that point regardless of coordinate system.

Equation \eqref{effmetric} may then be used to define an effective metric even when considering a quantum superposition of spacetimes by first computing the quantum average of the Wightman function and then using \eqref{effmetric}. Thus \eqref{effmetric} may regarded as an improved field theory expression for the average metric \eqref{metricave} which accounts for the actual propagating states of the system. From the perspective of our current discussion, \eqref{effmetric} is still just a single metric with its own causal structure which is necessarily additive, and so \eqref{effmetric} cannot account for the failure of additivity. It would however be interesting to see if an equivalent of $C$ could be defined directly in terms of correlation functions.

\subsection{Two light cones}

Let us then take as a working hypothesis that $\bar d(x,y)$ gives an accurate description of causality. Related discussions of how causality may be imposed in theories where the spacetime geometry is fluctuating is given in \cite{Hardy:2005fq}. Qualitatively, we are assuming that ($a$) if a photon is sent at $x$, its expected time of arrival $y^0$ on the $\vec y$ wordline is well approximated by $\bar d(x,y)=0$, and ($b$) the average of commutators of (suitably dressed) local field operators  over the metric fluctuations $\langle [A(x),B(y)]\rangle$ drops off somewhere close to the $\bar d(x,y)=0$ surface. We also assume here that $\bar d(x,y)$ is \emph{subadditive}

Two geometric structures are at play. The first one is given by the average distance  $\bar d(x,y)$. The double derivative of  $\bar d(x,y)$ also allows to calculate a metric tensor $\bar g_{\mu \nu} (x)$ at every point through eq.~\eqref{metrictensor}. However,  $\bar d(x,y)$ \emph{is not} the geodesic distance associated with $\bar g_{\mu \nu} (x)$. So $\bar d(x,y)$ is more fundamental than $\bar g_{\mu \nu} (x)$, because we can derive the latter from the former but not \emph{vice-versa}. The average metric tensor $\bar g_{\mu \nu} (x)$ defines a pseudo-Riemannian geometry with its usual ``rigid" causal structure. Every event $x$  is the vertex of the light cone defined by $\bar g_{\mu \nu} (x)$. The latter approximates the behavior of $\bar d(x,y)$ \emph{locally}. It gives information about \emph{e.g.} experiments in which a photon is emitted at $x$ and received in the immediate vicinity. If the photon is received further away we cannot trust the rigid light cone of $\bar g_{\mu \nu} (x)$ anymore and we need to look at the condition 
$\bar d(x,y)=0$.
\begin{figure}[h]
%\vspace{-1cm}
\begin{center}
   \includegraphics[width=10cm]{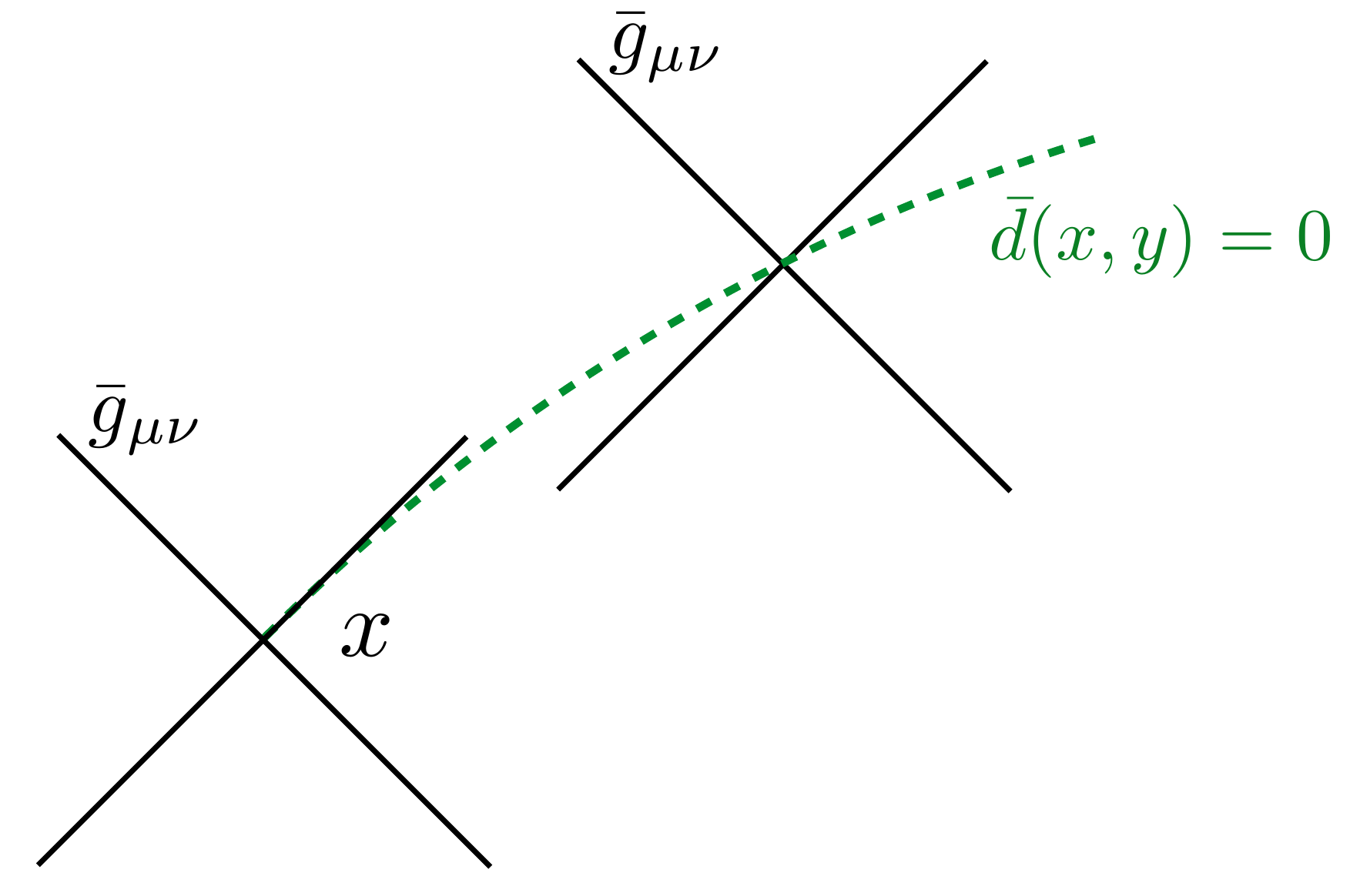}
  \end{center}
\caption{For any sufficiently smooth average distance $\bar d(x,y)$ one can define at any point an average metric tensor $\bar g_{\mu \nu}(x)$ through~\eqref{metrictensor}. If $\bar d(x,y)$ is \emph{subadditive}, the surface $\bar d(x,y)=0$  (dashed, in green) grows \emph{space-like} with respect to $\bar g_{\mu \nu}(x)$.} \label{fig-1}
  \end{figure}

For $x$ fixed, the surface defined by the points $y$ satisfying $\bar d(x,y)=0$ \emph{is not null} for the metric $\bar g_{\mu \nu}$.
This can be seen directly by the subadditivity condition, 
\begin{equation} \label{crucial-end}
C(x,y) = \frac14\,  \frac{\partial \,  {\bar d}^2(x,y)}{\partial y^\mu} \frac{\partial \, \bar d^2(x,y)}{\partial y^\nu} \ \bar g^{\mu \nu}(y) - \bar d^2(x,y)\, <0 .
\end{equation}
The above inequality implies that, on $\bar d(x,y)=0$, the gradient of $\bar d(x,y)$ itself is \emph{timelike}---\emph{i.e.} the surface $\bar d(x,y)=0$ is space-like---with respect to the metric $\bar g_{\mu \nu}$. This means that, at large separation, the rigid light cone structure associated with $\bar g_{\mu \nu}$ \emph{overestimates} the expected time of arrival of the photon (Fig.~\ref{fig-1}). In this sense, the photon is \emph{prompt} with respect to the classical expectation suggested by $\bar g_{\mu \nu}$.

Notice that the surface $\bar d(x,y)=0$ (for fixed $x$) has the same topology as a standard light cone but it is not ``rigid", in the sense that, in the vicinity of some point $y$, is not characterized by anything locally defined at $y$. The inclination of $\bar d(x,y)=0$ at $y$ depends on where the other extreme $x$ is located. Also, $\bar d(x,y)=0$ is not a collection of light rays  and does not contain the ``trajectory" of any photon. Similarly to the standard double slit experiment, we only have a handle on the expected time of arrival of the photon---this is what $\bar d(x,y)=0$ keeps track of. Before its detection at $y$, the photon lives on a superposition of geodesics and it is not associated with any particular trajectory.

\subsection*{Note on superadditivity}
 By inverting the inequality in~\eqref{crucial-end} one finds that, if $\bar d(x,y)$ is superadditive, $\bar d(x,y)=0$ is a \emph{timelike} surface with respect to $\bar g_{\mu \nu}$. While $C<0$ makes the domain of causal influence of some event $x$ more inclusive than the standard one, the $C>0$ case tends to exclude points along the way. By defining causal relations $x\prec y$ with $\bar d(x,y)$ (i.e. $\bar d(x,y)^2<0$ and $y$ in the future of $x$) it is not difficult to see that superadditivity violates \emph{transitivity}, i.e.
\begin{equation}
x\prec y \ \wedge \ y \prec z \ \ \nRightarrow \ \ x\prec z\, .
\end{equation}
If the average of local commutators drops off across $\bar d(x,y)=0$ relatively sharply, this corresponds to the following situation
\begin{equation}
\bra[{\cal A}(x), {\cal A}(y)]\ket\neq 0, \quad \bra[{\cal A}(y), {\cal A}(z)]\ket\neq 0, \quad \bra[{\cal A}(x), {\cal A}(z)]\ket \approx 0\, ,
\end{equation}
where ${\cal A}(x)$ represents the algebra of local operators at $x$. These relations among the local algebras of three events are never realized in the standard case (say in a QFT in Minkoswki). By contrast, subadditivity does not seem to lead to relations among local algebras  that are not already contemplated in standard QFT.

\subsection{A conjecture}

We have shown the Euclidean distances are always subadditive. For Lorentzian, we have collecting evidence in favor of subadditive distances, i.e. $C\leq0$ as being an appropriate physical requirement. The example of Sec.~\ref{FRW} shows that it is possible to build superadditive Loretnzian distances, by choosing a superposition of FRW spacetimes with equation of state sufficiently negative. In this example, however, the fluctuations of the spatial derivatives of the metric, which give a negative contribution to $C$ (see~\eqref{3.11}-\eqref{3.14}), are turned off by hand from the outset. This is obviously an artifact of the brutal mini-superspace approximation (see also~\cite{Nicolis:2022gzh} on this). If spatial derivatives weight as much as time derivatives in the expression for $C$ then the negative term~\eqref{3.14} would overrule the positive contribution~\eqref{3.11} in generic situations. This is what happens in all the other examples that we have considered. Naively, subadditivity seems to follow from the fact that ``space wins over time" as number of dimensions.
On top of this, superadditivity is problematic from the point of view of causality because, as just seen, it violates the transitivity of causal relations. 

The FRW example shows that superadditive distances can be engineered.  However, in any discussion of causality for standard Lorentzian geometries, it is important to ask questions related to reasonable physical initial data. For example, we are not concerned whether a given theory admits solutions with closed timelike curves, we are only concerned if they can form from reasonable initial data. Similarly, here, we are interested in situations where large metric fluctuations---and sizable non-additivity---develop as the result of dynamical evolution. The spreading of the wavefunction of the metric could happen through e.g. some tunneling-like event for the gravitational field. 
It is thus tempting to conjecture that \emph{subadditivity is preserved by time evolution and/or develops from reasonably generic initial conditions.} 
This may be regarded as a variant of the chronology protection conjecture \cite{Hawking:1991nk}: subadditivity begets subadditivity.

\section{Discussion}

The perspective taken in this article is that the failure of average distances to be additive can be a useful diagnostic to probe fluctuating geometries, both when the fluctuations are small such as vacuum or thermal fluctuations, or when they are large when for instance the state of the system me be regarded as a superposition of semi-classical geometries. We have shown that averages of Euclidean geometries are always subadditive, and they respect the triangle inequality. In this sense the average squared distance can be used to define an effective metric space.

The Lorentzian case is notably richer, and while we cannot prove subadditivity at present, it is a property of many physical situations. Furthermore, the violation of transitivity of causal relations implied by superadditivity is strongly suggestive that real physical systems with meaningful initial data respect subadditivity.  One of our central points is that the `metric is not enough', that is the quantum geometry is better described by an average distance than by an average metric, since the latter cannot account for the property of subadditivity. The absence of a fixed metric, or clear rigid causal structure brings up many interesting conceptual questions.

{\bf Apparent Superluminality:} Inspecting Fig.~\ref{fig-1} gives the appearance that something propagating along the curve $\bar d(x,y)=0$ propagates superluminally. However, this word is misleading in the context of \emph{subadditive distances}, because nothing physical travels along the light cone of the local average metric $\bar g$. One way to trace the rigid light cone of $\bar g_{\mu \nu}$ would be to engineer multiple experiments  where photons are emitted and immediately detected all along the light cone (see e.g. Fig. 4 of~\cite{Piazza:2021ojr}). (Curiously, the very same experiment would allow a faster transfer of information if $C>0$, which is another puzzling and possibly unacceptable feature of superadditive distances.)  For the same reasons, there are no violations of causality here. However, if one mistakenly regarded $\bar g_{\mu \nu}$ as all there is to know about the system, their \emph{classical expectations about causality} might indeed be violated. In other words the causal structure of a quantum geometry is not determined by its average metric.

{$\boldsymbol{\bra d \ket \ \rm{vs.} \ \sqrt{\bra d^2\ket}}$:} One may wonder if nonadditivity is an artifact of having chosen $\bar d(x,y) = \sqrt{\bra d^2(x,y)\ket}$ as the quantity to look at and not simply $\bra d(x,y)\ket$. As already pointed out, $\bra d(x,y)\ket$ is problematic to define in Lorentzian signature, and it is for that reason that the Hadamard structure of propagators in curved spacetime is determined by an analytic function of $d^2$, not of $d$. At the same time, $\bar d(x,y)$ has the correct \emph{local} behavior of a distance. From $\bar d(x,y)$ we can define a local metric tensor $\bar g_{\mu \nu}(x)$ through eq.~\eqref{metrictensor}. But then a simple coordinate expansion~\eqref{dsquare2} shows that $\bar g_{\mu \nu}(x)$ 
is nothing else than $\bra g_{\mu \nu}\ket$, the average metric tensor in $x$. If we did the same exercise with $\bra d(x,y)\ket$ there would not be such a nice identification and the information about $\bra g_{\mu \nu}\ket$ would be lost. At any rate, it is not difficult to show that  $\bra d(x,y)\ket$---say, in Euclidean signature and in more than one dimension---is also non additive in general. For a classical ensemble the source of nonadditivity is the fact that the average can happen over different geodesics taking different paths on each classical representative. 

{\bf Relativity of the event:} The prescription for defining an event on a superposition of spacetimes is by no means unique. In asymptotically AdS one can use \emph{spacelike} geodesics to anchor bulk points to the boundary. Timelike geodesics---used in this paper---are perhaps more physical because they can disclose the actual experiences of a bunch of observers. Still, different \emph{classes of observers} will generally define what is an event in different ways. Consider a theory with \emph{two} non relativistic fluids, $A$ and $B$, 
%\begin{equation} \label{2-conclusion} \nonumber
%S  \ = \  \frac{1}{16 \pi G}\int d^4 x \sqrt{-g} R \   - \ \mu_A^4 \int d^4 x \sqrt{-g} \sqrt{\det g^{\mu \nu} \partial_{\mu }X^i_A \partial_{\nu}X^j_A}\   - \ \mu_B^4 \int d^4 x \sqrt{-g} \sqrt{\det g^{\mu \nu} \partial_{\mu }X^i_B \partial_{\nu}X^j_B}+ \ \dots\, .
%\end{equation}
\begin{equation} \label{2-conclusion} \nonumber
S  \ = \  - \ \mu_A^4 \int d^4 x \sqrt{-g} \sqrt{\det g^{\mu \nu} \partial_{\mu }X^i_A \partial_{\nu}X^j_A}\   - \ \mu_B^4 \int d^4 x \sqrt{-g} \sqrt{\det g^{\mu \nu} \partial_{\mu }X^i_B \partial_{\nu}X^j_B}+ \ \dots\, .
\end{equation}
For metric superpositions that share a common classical Minkowski space at early times (e.g. the situation discussed in the introduction and App.~\ref{grav-waves}), a neat example would be to make the $B$-observers initially boosted with respect to the $A$-observers by some constant velocity. 

If we adopt unitary gauge for the observers $A$, non additivity disappears along the $A$-geodesics, as discussed in Sec.~\ref{subec-lorentzian}.
However, in these $x_A$-coordinates, the trajectories of the $B$-fluid will contain some uncertainty because of the fluctuations in the gravitational field. But the $B$-observers define the events $x_B$ with \emph{their} trajectories. This means that, as opposed to classical general relativity, there is no one to one correspondence of the type $x_A = x_A(x_B)$ between the two coordinate systems $x_A$ and $x_B$. There is instead a \emph{probability amplitude} that some event $x_B$ coincides with some event $x_A$. In the $x_A$ coordinate system, an event $x_B$ is spread over different locations, each weighted by some probability. In the context of JT-gravity this frame ambiguity has been highlighted in~\cite{Blommaert:2019hjr}.

{\bf Quantifying the effect:} Non additivity is quantified by $C(x,y)$ defined at eq.~\eqref{crucial-intro}. This quantity has dimensions of a length squared and grows with the separation between $x$ and $y$.  A way to quantify the effect is to estimate at which distance $\bar d(x,y)$ we have $C(x,y)/\bar d^2(x,y) \sim 1$. When considering $x$ and $y$ at null separation one should refer to a characteristic length scale $\ell$ in the problem. The larger is $\ell$, the smaller the effect.  We have calculated $C$ in simple perturbative situations. A thermal state of gravitons at temperature $T$, for instance, breaks Lorentz invariance, so that one can associate a transverse length $\ell$ to any point along the light cone. We calculated $C$ in App.~\ref{app-thermal} finding that non-additivity becomes sizable at lengths of order $\ell \sim M_{Pl}/T^2$ (see App.~\ref{app-thermal}). If gravitons were in equilibrium during radiation domination this would be the size of the Hubble horizon. However, non-additivity could be enhanced if the wavefunction of the metric is very ``spread". Non-perturbative mechanisms could be at play that make the state of the gravitational field a coherent superposition of macroscopically different configurations. This is what happens for the $\theta$-vacua of Yang-Mills theories, which are superpositions of different winding numbers. 

{\bf Quantum coherence:}  In order to have the claimed subadditivity we need to maintain some quantum coherence in the gravitational field, which may be lost as soon as we measure something (say, $H_0$). 
 Any observation will clearly decohere the system to some extent. However, it would be important to understand on which \emph{pointer basis} the system decoheres into.  No realistic measurement can meaningfully collapse a superposition of semi-classical states into a single semi-classical geometry for which the effects discussed in this paper would disappear altogether after the first observation. Indeed in the real universe, we `measure' the geometry of spacetime by means of standard candles and standard rulers which themselves depend on distances. The effective geometry is inferred from a statistical distribution of distance and redshift measurements.  According to the textbook interpretation, the system should collapse into an eigenstate of the observable, which in this case probably makes a very highly degenerate subspace. This is clearly a point that deserves better understanding. We note, also, that the most promising directions towards the solution of the black hole information paradox are truly quantum informational in nature~\cite{Hayden:2007cs,Yoshida:2017non}, and require total coherence of the system until the Hawking radiation is collected at infinity. 

{\bf Spontaneous Symmetry breaking:} Related to the above point is that in a field theory some coherent superpositions are effectively forbidden and one is immediately projected in one of the (possibly infinite) superselection sectors. Whether this happens or not in dynamical gravity is questionable, however. On the one hand, there are indications that the total number of degrees of freedom might be finite (e.g. limited by the exponential of the Beckenstein-Hawking entropy in the case of a black hole). Also, spontaneous symmetry breaking does not apply to those situations where the Euclidean action that interpolates the different classical vacua is finite. Again, the vacuum of Yang-Mills theories is an example of a genuine coherent superposition of different semiclassical configurations. 

\bigskip
\noindent{\textbf{Acknowledgments:}} We thank Alberto Nicolis for insightful exchanges and Alexander Taskov for initial collaboration.  We also  thank the organizers and participants of the Simons Symposium on Origins of the Universe, held in April 2022, for interesting, nae inspiring discussions. The work of FP is supported by the Programme National GRAM of CNRS/INSU with INP and IN2P3 co-funded by CNES. The work of AJT is supported by STFC grant ST/T000791/1. AJT thanks the Royal Society for support at ICL through a Wolfson Research Merit Award.

\appendix
%\addtocontents{toc}{\protect\setcounter{tocdepth}{1}}

\section{Some results about the fluid of observers} \label{app-fluid}

 \subsection{The observers are free-falling}
 Let us check explicitly that $x^i =$ constant is a geodesic on any given classical solution of action~\eqref{2}. In order for this to be the case we should have $\Gamma^i_{00} = 0$. By contracting this condition with $g_{\mu i}$ we can re-write it as
\begin{equation}
g_{\mu \nu} \Gamma^\nu_{00} - g_{\mu 0} \Gamma^0_{00}=0\, .
\end{equation}
  The zero component of the above equation gives 
  \begin{equation}
  \frac12 g_{00,0} = g_{00} \Gamma^0_{00}, 
   \end{equation}
which allows to write the $i$ component condition as
\begin{equation} \label{eombyby}
g_{00} g_{0i,0} - \frac12 g_{00} g_{00,i} - \frac12 g_{0i} g_{00,0} = 0\, .
\end{equation}

We want to show that~\eqref{eombyby} is equivalent the eoms for $x^i$ derived from~\eqref{2} and written in unitary gauge. 
The eom for $x^i$ read
\begin{equation} \label{eombutso}
\partial_\mu\left[\sqrt{-g} (\det B^{IJ})^{1/2} B_{IJ}^{-1} g^{\mu \nu} \partial_\nu x^I\right] = 0.
\end{equation}
In unitary gauge $B^{IJ} = \gamma^{ij} - \frac{N^i N^j}{N^2}$, where we have used the common notation of the ADM formalism,
\begin{equation}
  g_{\mu \nu} = \begin{pmatrix} -N^{2} + N^{k}N_{k} \ \ \ & N_{j}\\[3mm]
                            N_{i} & \gamma_{i j}
                 \end{pmatrix} ,\qquad
  g^{\mu \nu} = \begin{pmatrix} - \dfrac{1}{N^2} \ \ \ \ \ \ &\dfrac{N^j}{N^2} \\[4mm]
                \dfrac{N^i}{N^2} \ \ \ &  \gamma^{i j}-\dfrac{N^i N^j }{N^2}
                \end{pmatrix} 
\end{equation}

 We also have 
\begin{equation} 
\det B^{IJ} = \gamma^{-1} \left(1 - \frac{N_k N^k}{N^2}\right).
\end{equation}
The fact that the square root of $\det B^{IJ}$ appears in~\eqref{2} allows an important simplification which allows to write~\eqref{eombutso}
as
\begin{equation}
\partial_0 \left[(-g_{00})^{-1/2} g_{0i} \right] + \partial_i (-g_{00})^{-1/2} = 0\, .
\end{equation}
The above condition is identical to~\eqref{eombyby}.

\subsection{The condition of zero shift}

Unitary gauge of the fluid is defined\footnote{With a mechanism well known also from particle physics, when the unitary gauge is adopted the fluid degrees of freedom are transferred to the metric. Here, in particular, to the three shifts $N^i$. The dynamics is exactly the same and the possible (tachyonic) instabilities appearing in the theory are the standard (Jeans-) ones of a fluid under its own gravity.} by fixing the 3 scalar functions $x^I = X^i$. This still leaves us the freedom of choice of the time coordinates. As stated in the introduction, we can fix the time by the proper time along each geodesic. In Unitary gauge, the proper time along a given geodesic is simply $\tau = \int  N d t$. Thus if we further choose time coordinates so that $N=1$, then $\tau = t$ and the equation for the fluid is
\be
\partial_0 \left[ \frac{N_i}{\sqrt{1-N_kN^k}}\right] =- \partial_i \left[\frac{1}{\sqrt{1-N_k N^k}}\right]  \, . 
\ee
Suppose now that on the initial surface $N^i=0$. The above equation then implies $\partial_0 N^i=0$ meaning that the shift remains zero.

It is easy to see that this is consistent with the gravitational equations of motion. In unitary gauge, the contribution of the perfect fluid to the gravitational equations is given by the fluid action:
\begin{equation}
S_{fluid} = - \mu^4 \int d^4 x N  \left(1 - \gamma_{i j} \frac{N^i N^j}{N^2}\right)^{1/2}\, .
\end{equation}
Therefore 
\ba
 \frac{\delta S_{fluid}}{\delta \gamma_{ij}} \quad \propto \quad N^i N^j \quad \xrightarrow[N^i \rightarrow 0]{} 0. \\
  \frac{\delta S_{fluid}}{\delta N^i}  \quad \propto  \quad N_i   \xrightarrow[N^i \rightarrow 0]{} 0.
\ea
This means that the dynamical equations for the metric and the usual momentum constraint are unaffected by the fluid. The only equation that is corrected is the Hamiltonian constraint by the addition of a constant
\be
 \frac{\delta S_{fluid}}{\delta N} = - \mu^4 \, .
\ee
which is just the statement that the fluid is pressureless and only contributes to the $G_{00}$ Einstein equation (Hamiltonain constraint). In order to neglect backreaction we require
\be
\mu^4 \ll M_{\rm Planck}^2 r_0^{-2}
\ee
where $r_0$ is the typical curvature length scale $R \sim 1/r_0^2$. This is equivalent to demanding that the curvature scale induced by the fluid $r_{\mu}=M_{\rm Planck}/\mu^2$ is much larger that than the curvature scales of interest $r_{\mu} \gg r_0$. As such we may neglect effects from $r_{\mu}$ in the computation of $d(x,y)$. At the same time, to safely trust the fluid effective theory we need to consider curvature scales below the EFT cutoff $1/r_0^2 \ll \mu^2$. Together these imply
\be
1/r_0^2 \ll \mu^2 \ll M_{\rm Planck} r_0^{-1} \rightarrow r_0 \gg M_{\rm Planck}^{-1} \, , 
\ee
which is easily satisfied as long as the spacetime curvature is well below the Planck scale. These conditions justify our neglect of the backreaction of the dilute gas.

\section{Spacetime additivity in arbitrary signature} \label{app-resum}

In any given normal neighborhood, we can write the distance from the ``origin" simply as $d^2 = Y^a Y^b \eta_{ab}$, where $Y^a$ are Riemann normal coordinates (RNC) and $\eta$ is the flat metric of Minkowski or Euclidean space. In any other coordinate system $x^\mu$ we want to evaluate
\begin{equation} \label{crucial}
C(0,x) = \frac14 g^{\mu \nu}(x) \partial_\mu d^2(0,x)\partial_\nu d^2(0,x)) - d^2(0,x)\, .
\end{equation} 
It is useful to understand first why $C$ vanishes in the standard case, when indeed $d^2 = Y^a Y^b \eta_{ab}$. We can write
\begin{equation}
C = g^{\mu \nu}(x) \partial_\mu Y^a \partial_\nu Y^c Y^b Y^d \eta_{ab} \eta_{cd}- Y^a Y^b \eta_{ab}= (g^{cd}(Y) \eta_{ac} \eta_{bd} -\eta_{ab}) Y^a Y^b\, ,
\end{equation}
where $g_{ab}(Y) = \eta_{ab} - \frac13 R_{acbd} Y^c Y^d + \dots$ is the metric in RNC. Clearly, $g_{ab}(Y)$, $\eta_{ab}$ and $g^{cd}(Y) \eta_{ac} \eta_{bd}$ are all  different quantities. However, they all behave identically when contracted with $Y$, because at any order in the RNC expansion of $g_{ab}(Y)$, the Riemann tensor appears already contracted twice with $Y$.  We conclude that 
\begin{equation} \label{conclude_app}
g^{\mu \nu}(x) \partial_\mu Y^a \partial_\nu Y^c Y^b Y^d \eta_{ab} \eta_{cd} = Y^a Y^b \eta_{ab},
\end{equation}
which, again, is nothing else than the additivity property that we have proven in various ways. 

RNC clearly have a privileged relation---even in curved space---with the flat metric $\eta$. We can exploit this further by introducing a tetrad/vierbein field $e^a_\mu$,  such that 
\begin{equation} \label{tetrads}
\eta_{ab} e^a_\mu e^b_\nu = g_{\mu \nu}, \quad g^{\mu \nu} e_\mu^a e_\nu ^b = \eta^{ab} \, .
\end{equation}
This allows to take the ``square root" of ~\eqref{conclude_app} and notice that $\partial_\mu Y^a$, despite being a different object than $e_\mu^a $,  behaves in the same way  when acting on the $Y$s, 
\begin{equation} \label{squared}
\partial_\mu Y^a \eta_{ab} \ Y^b \ = \ e^a _\mu \eta_{ab} \ Y^b .
\end{equation}

Before moving on to average metrics and distances let us introduce a more compact notation in which all indexes contractions become standard matrix multiplications. We thus abolish raising and lowering of indexes but we need to agree on a convention for their ordering. We write
\begin{equation}
Y^a \rightarrow v\, , \quad g_{\mu \nu} \rightarrow g \, , \quad e^a_\mu \equiv (e)_{ a\mu} \rightarrow e \, .
\end{equation}
Eqs.~\eqref{tetrads} become 
\begin{equation} 
g = e^T \eta e, \qquad \eta = e \, g^{-1} e^T\, .
\end{equation}
Keeping~\eqref{squared} into account, eq. \eqref{conclude_app} becomes
\begin{equation}
v^T \eta\, e\  g^{-1} e^T \eta \, v \ = v^T \eta \, v\, .
\end{equation}

\subsection{Statistical mixtures}
Each element of a statistical mixture of metrics can be defined by the relation that RNCs $Y^a_n$ have with the physical coordinate system of observers $x$. By labeling each element of the statistical ensemble with the letter $n$ we write the average distance as
\begin{equation}
{\bar d}^2 = \sum_n p_n\ y^{a}_{n} \eta_{ab} y^b_{n} = \sum_n p_n \ v^T_n \eta\,  v_n\, , 
\end{equation}
with $\sum_n p_n = 1$.
In compact notation we simply call $ g$ the average of the metrics, and $ g^{-1}$  its inverse.

Let us consider the mixture of only two metrics, $g_0$ and $g_1$ to start with,
\begin{align} 
g &= (1-p) g_0 + p g_1 \\ \label{mixture}
& = (1-p) g_0 \left(1+ \frac{p}{1-p} g_0^{-1} g_1\right) \, .
\end{align}
Each metric comes with their tetrads, $g_0 = e_0^T \eta e_0$, $g_1 = e_1^T \eta e_1$. Each metric produces a distance that satisfies the additivity property $C=0$. In particular, 
\begin{equation} \label{virtue}
v_0^T \eta \, e_0\  g_0^{-1} e_0^T \eta\,  v_0 \ = v_0^T \eta \, v_0\, .
\end{equation}

The expression in~\eqref{mixture} suggests the use of the expansion parameter 
\begin{equation}
b \equiv \frac{p}{1-p} \,
\end{equation}
In terms of which the inverse metric can be expressed with a power series, 
\begin{equation}
p \, g^{-1} = \sum (-1)^n b^{(n+1)} \left(g_0^{-1} g_1 \right)^n g_0^{-1}\, .
\end{equation}
The spacetime additivity~\eqref{crucial} then reads
\begin{align} \label{rearrange}
C & = \left[(1-p) v_0^T \eta e_0 + p v_1^T \eta e_1\right] g^{-1} \left[(1-p) e_0^T\eta  v_0 + p e_1^T \eta v^T\right] - (1-p)\,  v_0^T\eta  v_0 - p\, v_1^T \eta v_1\\ \nonumber
 = &\, p \left\{ \left[\frac{ v_0^T \eta e_0}{b} + v_1^T \eta e_1\right] \sum_{n=0}^\infty (-1)^n b^{(n+1)} ( g_0^{-1} g_1)^n g_0^{-1} \left[\frac{A_0^T \eta v_0}{b} + e_1^T \eta v_1\right] -\frac{v_0^T\eta  v_0}{b} - v_1^T \eta v_1 \right\}\, .
\end{align}
In this complicate expression we can recognize that the leading order in $b$ [i.e. ${\cal O}(b^{-1})]$ cancels in virtue of~\eqref{virtue}. Then by writing out the first few terms one sees that the entire expression can be put in the following form
\begin{equation} \label{leqleq}
C = - p\,  V^T \eta \, V\, 
\end{equation}
where 
\begin{equation}
V = \frac{1}{\sqrt{1-b \left(A_1 g_0^{-1} e_1^T\eta \right)} }\left(A_1 g_0^{-1} e_0^T \eta \,  v_0 - v_1\right)\, .
\end{equation}
When working with Euclidean signature $\eta_{ab} = \delta_{ab}$ and from~\eqref{leqleq} we obtain $C\leq 0$ by construction:
\be
C= - p \sum_a (V^a)^2 \le 0 \, .
\ee 

\section{Triangle inequality of average distances} \label{app-triangle}

Consider a situation in which the distances among  a group of elements are given statistically. Two elements $x$ and $y$ are at distance $d_i(x,y)$ with probability $p_i$, $\sum^N_i p_i = 1$. The average distance between $x$ and $y$ is $\langle d(x,y)\rangle = \sum_i p_i d_i(x,y)$.
We assume that the triangle inequality is satisfied separately in each element of the ensemble. So if we take three elements $x$, $y$ and $z$ we assume that, for each $i=1\dots N$, 
\begin{equation}
d_i(x,z) + d_i(z,y) \geq d_i(x,y). 
\end{equation}
It follows by linearity that the triangle inequality is also satisfied by average distances, 
\begin{equation}
\langle d(x,z)\rangle + \langle d(z,y) \rangle \geq \langle d(x,y) \rangle. 
\end{equation}
Here we want to prove that the triangle inequality is also satisfied at the level of the \emph{root mean square} distance, 
\begin{equation}
\bar d(x,y) \equiv \sqrt{\langle d(x,y)^2\rangle} = \sqrt{\sum_i^N p_i d_i(x,y)^2}\, .
\end{equation}
In order to simplify the notation let us use 
$A_i \equiv d_i(x,z)$, $B_i\equiv d_i(z,y) $, $C_i\equiv d_i(x,y)$ from now on.
We want to show that 
\begin{equation} \label{triangle}
A_i + B_i \geq C_i
\end{equation}
implies
\begin{equation} \label{a.4}
\bar A + \bar B  \geq \bar C. 
\end{equation}

To start with, we show that~\eqref{a.4} is satisfied when the ensemble is made by only two elements, of probabilities $p$ and $1-p$ respectively. In this case, 
\begin{align}
&\ \bar A + \bar B  - \bar C \label{initial}\\ \equiv &\ \sqrt{p A^2_1 +(1-p) A_2^2} \, + \sqrt{p B^2_1 +(1-p) B_2^2}\, - \sqrt{p C^2_1 +(1-p) C_2^2} \\ 
\geq&\ \sqrt{p A^2_1 +(1-p) A_2^2} \, + \sqrt{p B^2_1 +(1-p) B_2^2}\, - \sqrt{p (A_1 + B_1)^2 +(1-p) (A_2 + B_2)^2} \label{final}.
\end{align}
The last step has been obtain by saturating the triangle inequality~\eqref{triangle} for both elements in the ensemble. We want to show that the expression at line~\eqref{final} must be positive, from which~\eqref{a.4} would follow. 
In order to show this we show that the opposite is a contradiction. Indeed, if the expression at~\eqref{final} were negative we would have 
\begin{equation}
\sqrt{p A^2_1 +(1-p) A_2^2} + \sqrt{p B^2_1 +(1-p) B_2^2} < \sqrt{p (A_1 + B_1)^2 +(1-p) (A_2 + B_2)^2} \, .
\end{equation}
By squaring both sides and making some simplifications we deduce 
\begin{equation}
\sqrt{p A^2_1 +(1-p) A_2^2}  \sqrt{p B^2_1 +(1-p) B_2^2} < p A_1 B_1 + (1-p) A_2 B_2\, .
\end{equation}
We square both sides again to finally obtain
\begin{equation}
p(1-p) \left(A_2 B_1 - A_1 B_2\right)^2 <0,
\end{equation}
which is clearly impossible. We deduce that the expression at line~\eqref{final}, and thus that at line~\eqref{initial}, is positive.

The full demonstration for a statistical ensemble of $N$ elements follows by reiteration. At each step we use the idea that the average of $n+1$ elements can be split into the average of the first $n$ elements and the last element. More formally, we can define the \emph{bar quantities} where averages are taken only among the first $n$ elements, 
\begin{equation}
\bar A_{(n)}^2 \equiv \frac{1}{\sum_i^n p_i} \sum _i^n p_i A_i^2\, ,
\end{equation}
and similarly for $B$ and $C$. Clearly we have $\bar A = \bar A_{(N)}$. Now it is easy to show that from 
\begin{equation} \label{assumption-reit}
\bar A_{(n)} + \bar B_{(n)} \geq \bar C_{(n)} 
\end{equation}
and~\eqref{triangle} it follows that 
\begin{equation} 
\bar A_{(n+1)} + \bar B_{(n+1)} \geq \bar C_{(n+1)} \, .
\end{equation}

Let us consider the $A$ term in the above equation, 
\begin{equation}
\bar A_{(n+1)} ^2 = \frac{1}{\sum_i^{n+1} p_i} \sum _i^{n+1} p_i A_i^2\ = \ p \bar A_{(n)}^2 + (1-p) A_{n+1}^2\, ,
\end{equation}
where 
\begin{equation}
p = \frac{\sum_i^{n} p_i}{\sum_i^{n+1} p_i}\, .
\end{equation}
So the problem is now identical to the bar-average of two elements, eqs.~\eqref{initial}-\eqref{final}, 
\begin{align}
&\ \bar A_{(n+1)} + \bar B_{(n+1)}   - \bar C_{(n+1)}  \\ = &\ \sqrt{p \bar A_{(n)}^2 + (1-p) A_{n+1}^2} \, + \sqrt{p \bar B_{(n)}^2 + (1-p) B_{n+1}^2}\, - \sqrt{p \bar C_{(n)}^2 + (1-p) C_{n+1}^2} \\ 
\geq&\ \sqrt{p A^2_1 +(1-p) A_2^2} \, + \sqrt{p B^2_1 +(1-p) B_2^2}\, - \sqrt{p (\bar A_{(n)} + \bar B_{(n)})^2 +(1-p) (A_{n+1} + B_{n+1})^2} .\nonumber
\end{align}
This time, in the last line, assumption~\eqref{assumption-reit} has been used together with~\eqref{triangle}. The rest follows like in the two-elements case previously discussed. 

\section{Gravitational waves: numerical analysis} \label{grav-waves}

\subsection{Two classical solutions with opposite polarization}

A useful reference for this section is~\cite{Zhang:2017geq}. We start by writing a planar wave vacuum solution in ``Brinkmann coordinates". As those are not observers' coordinates we denote them with capital letters. 
\begin{equation} 
ds^2 = - 2 dU dV + dX^2 + dY^2 + \left[{\cal A}_+(U) (X^2 - Y^2) + 2 {\cal A}_\times (U)\, X Y\right] dU^2,
\end{equation}
where ${\cal A}_+$ and ${\cal A}_\times$ are generic functions of $U$. Let us consider directly the two simple metrics that we are going to study in  detail. The \emph{plus} metric has ${\cal A}_+(U) = \theta(U) A_+$, ${\cal A}_\times (U)=0$. The \emph{cross} metric has  ${\cal A}_+(U) = 0$, ${\cal A}_\times (U)= \theta(U) \At$,
\begin{align}
ds_+^2 &= - 2 dU dV + dX^2 + dY^2 +  A_+ (X^2 - Y^2) \theta (U) dU^2,\\
ds_\times^2 &= - 2 dU dV + dX^2 + dY^2 +  2   A_\times \, X \, Y \ \theta(U) dU^2.
\end{align}
Both metrics are initially Minkowski with a gravitational wave incoming from the negative $Z$ axis. We can imagine a continuum of free falling test particles initially at rest in Minkowski space getting perturbed by the arrival of the wave.  We label these particles by their initial Minkowski coordinates and use their proper times as the time coordinate $t$. In order to do this, it is sufficient to write an appropriate set of geodesics for the two metrics. Explicitly,  for $U>0$,
\begin{align}
X&= x \cosh(\sqrt{A_+} U),  \\
Y&= y \cos(\sqrt{A_+} U),\\ 
V&= \frac{1}{\sqrt{2}}(t + z) + \frac12\left[x'(U) x(U) + y'(U) y(U)\right]  ,\\
U&=\frac{1}{\sqrt{2}}(t-z)\, .
\end{align} 
For $U<0$, simply, $X= x$, $Y=y$, $V=(t + z)/\sqrt{2}$, $U=(t - z)/\sqrt{2}$.  Of course, $x$, $y$, $z$ and $t$ can also be used as coordinates, with the metric taking the ``Baldwin-Jeffery-Rosen" form. At $U>0$, or $t>t$, the metric reads
\begin{equation} \label{rosen-1}
ds_+^2 = - dt^2  + \left[1 + \sinh^2\left(\sqrt{A_+}\, (t-z)\right)\right] dx^2 + \left[1 - \sin^2\left(\sqrt{A_+}\, (t-z)\right)\right] dy^2 + dz^2\, .\\
\end{equation}

Something completely analogous can be done for the \emph{cross} metric. The geodesics at $U>0$ read
\begin{align}
X&= \frac12 \left[(x-y) \cos(\sqrt{A_+} U) + (x+y) \cosh(\sqrt{A_+} U)\right], \\
Y&= \frac12 \left[(y-x) \cos(\sqrt{A_+} U) + (x+y) \cosh(\sqrt{A_+} U)\right],\\ 
V&= \frac{1}{\sqrt{2}}(t + z) + \frac12\left[x'(U) x(U) + y'(U) y(U)\right] ,\\
U&=\frac{1}{\sqrt{2}}(t-z)\, .
\end{align}
The metric in these coordinates, for $U>0$, becomes
\begin{align} \nonumber
ds_\times^2  = \ &- dt^2  + \tfrac12 \left[\cosh^2\left(\sqrt{A_+}\, (t-z)\right)+\cos^2\left(\sqrt{A_+}\, (t-z)\right) \right] (dx^2 + dy^2) \\[2mm]
&+ \left[\cosh^2\left(\sqrt{A_+}\, (t-z)\right)-\cos^2\left(\sqrt{A_+}\, (t-z)\right) \right] dx dy + dz^2\, .  \label{rosen-2}
\end{align}
Both~\eqref{rosen-1} and~\eqref{rosen-2} are of the form~\eqref{null-metric}. When taking quantum superpositions of them we can simply rely on the general results of Sec.~\ref{sec_planewaves}. The aim of this appendix, however, is to do things very explicitly.

\subsection{Distances}
For any classical solution~\eqref{rosen-1} or~\eqref{rosen-2} we can compute the distance between any two events $d(x_1, x_2)$. While no close expression for $d(x_1, x_2)$ seems available, Mathematica can easily expand this quantity in the parameter $A_{+\times}$. The results  are summarized below.

We checked that the long and rather complicated expressions of  $d^2(x_1, x_2)$ satisfy the additivity condition both for the \emph{plus} and the \emph{cross} metrics: given two points $x_1$ and $x_2$ with $d^2(x_1, x_2)=0$, and a separating hypersurface between them, the problem of finding a third point $x_3$ on the surface with both $d^2(x_1, x_3)=0$ and $d^2(x_3, x_2)=0$ \emph{has only one solution}. To order $A^2$ this amounts to solving a set of second order equations that happen to have exactly zero discriminant. 

The next step was to take \emph{averages} of distances. We have calculated\footnote{Notice that this corresponds to taking an average on a classical ensemble of density matrix
\be
\hat \rho = c  |\psi_1 \rangle \langle \psi_1|+ (1-c)  |\psi_2 \rangle \langle \psi_2|\, .
\ee
Or, to a situation where the cross terms $\langle \psi_1|d^2(x_1, x_2)|\psi_2 \rangle$  are negligible (this is the case when $\langle \psi_1|\psi_2 \rangle$ is itself negligeable---we thank Alexander Taskov for pointing this out). 
However, no such restriction is assumed in the formulae of Sec.~\ref{sec_normal}. } 
\begin{equation}
\bar d(x_1, x_2) ^2 = c \, d_1(x_1, x_2)^2  + (1-c)  d(x_1, x_2)^2 
\end{equation}
for different values $0<c<1$ and when $(a)$ the two classical distances $d_1$ and $d_2$ correspond to different polarizations, $(b)$ when they have the same polarizations but different intensities, e.g. $A'_+ \neq  A_+$. In all cases \emph{the third point problem}  for two initial points at zero distance does not have any real solution, as expected in the case of subadditivity. Again, to order $A^2$ this amounts to solving a set of second order equations which this time have \emph{negative} discriminant. The explicit expressions of the distances are too long to be displayed. But we have reproduced case $(b)$ in the simplified setup of scalar gravity where the expressions are more manageable.

\subsection{Scalar waves}

We now consider a scalar tensor theory of gravity in which matter is minimally coupled to the metric $g_{\mu \nu} = e^{2\phi} \eta_{\mu \nu}$, where $\phi$ is a scalar field. The Christoffel symbols in this case read
\begin{equation}
\Gamma^{\rho}_{\mu \nu} = \delta^\rho_\mu \partial_\nu \phi + \delta^\rho_\nu \partial_\mu \phi - \eta_{\mu \nu} \eta^{\rho \sigma}\partial_\sigma \phi\, .
\end{equation}
As before, we can imagine a plane wave traversing Minkowski space in the $z$ direction. We can also work directly in null coordinates, so that $\phi = f(U)$ and computations simplify considerably, 
\begin{equation}\nonumber
\Gamma^U_{\mu \nu} = 
\begin{pmatrix}
2 f'& 0 &0&0\\
0&0&0&0\\
0&0&0&0\\
0&0&0&0\\
\end{pmatrix}\!,
\Gamma^V_{\mu \nu} = 
\begin{pmatrix}
0& 0 &0&0\\
0&0&0&0\\
0&0&f'&0\\
0&0&0&f'\\
\end{pmatrix}\!,
\Gamma^X_{\mu \nu} = 
\begin{pmatrix}
0& 0 &f'&0\\
0&0&0&0\\
f'&0&0&0\\
0&0&0&0\\
\end{pmatrix}\!,
\Gamma^Y_{\mu \nu} = 
\begin{pmatrix}
0& 0 &0&f'\\
0&0&0&0\\
0&0&0&0\\
f'&0&0&0\\
\end{pmatrix}\!.
\end{equation}
In order to solve the geodesic equations explicitly we chose $f(U) = A\, U \, \theta(U)$. As before, spacetime is Minkowski before the arrival of the  wave. The geodesic equations read
\begin{align}
& U''(\lambda) + 2 A \, U'(\lambda) = 0\, ,\\
& V''(\lambda) + A (X'(\lambda)^2 + Y'(\lambda)^2) = 0\, ,\\
& X''(\lambda) + 2 A \, U'(\lambda) X'(\lambda) = 0\, ,\\
& Y''(\lambda) + 2 A \, Y'(\lambda) Y'(\lambda) = 0\, .
\end{align}
If one looks at points on  the $X=Y=0$ plane, all geodesics connecting them also belong to such a plane and the problem becomes effectively two dimensional. However, in this case the results become too trivial. Things are still interesting if we only restrict to $Y=0$. 

As before, we can populate the spacetime with free falling test particles with labels $x$, $y$, $z$ and proper time $t$. We can solve the geodesic equations and expand the distance with the aid of Mathematica. This time the expression is manageable, 
\begin{align} \label{5.6}
d(x^\mu_1, x^\mu_2)^2 =\ &d_0(x^\mu_1, x^\mu_2)^2 + \frac{(t_1 + t_2 + z_1 + z_2)(x_1 - x_2)^2}{\sqrt{2}} A \\
&- \frac16 \, (t_1 - t_2 + z_1 - z_2)^2\, (x_1 - x_2)^2 A^2 + \dots\, , \label{5.7}
\end{align}
where $d_0(x^\mu_1, x^\mu_2)$
is the Minkowski distance.  When superpositions are taken of different spacetimes  with different values of the $A$ parameter, averages of $A$ appear in~\eqref{5.6} and~\eqref{5.7}, 
\begin{align} \label{5.9}
\bar d(x^\mu_1, x^\mu_2)^2 =\ &d_0(x^\mu_1, x^\mu_2)^2 + \frac{(t_1 + t_2 + z_1 + z_2)(x_1 - x_2)^2}{\sqrt{2}} \langle A\rangle \\
&- \frac16 \, (t_1 - t_2 + z_1 - z_2)^2\, (x_1 - x_2)^2 \langle A^2 \rangle + \dots\, .  \label{5.10}
\end{align}
Notice that, to first order in $A$, or in the coordinate expansion, there is no difference between an actual spacetime and a average of spacetimes. The difference appears at second order as, in general, $\langle e^2 \rangle \neq \langle A \rangle^2$, and additivity is lost. 

Indeed, with Mathematica, we can test a different distance than~\eqref{5.6}-\eqref{5.7}, with the one sixth factor in the second line substituted by some more general coefficient, 
\begin{equation} -\frac16 \quad \rightarrow \quad - \frac{c}{6}\label{substitute}
\end{equation}
The ``third point" problem becomes then a set of quadratic equations, all with discriminant $\sqrt{1-c}$. While $c=1$ corresponds to the additive case,  for $c>1$ the third point problem has no real solutions, which we associate to \emph{subadditivity}. But that $c\geq1$ simply follows from the known inequality $\langle A^2 \rangle \geq \langle A \rangle^2$.

\section{Spacetime additivity of a thermal state of gravitons}\label{app-thermal}

In the case of gravitational perturbations around Minkowski we can evaluate $C$ in the vicinity of the light cone by using eq.~\eqref{c1} ,
\begin{align} \label{app-c1}
C  = & -\frac{t^4}{4}\left[- \frac14 \langle \dot h_{ij} \dot h_{kl} \rangle u^i u^j u^k u^l + \langle \dot h_{p i} \dot h_{p j}\rangle  u^i u^j \right. \\   \label{app-c2} 
&+ \left(2 \langle \dot h_{pk} h_{i p,j}\rangle -\langle \dot h_{pk} h_{i j,p}\rangle \right)  u^i u^j u^k\\  \label{app-c3}
&+ \left. \left(\langle h_{i p,j}h_{k p,l} \rangle - \langle h_{i p,j}h_{k l,p} \rangle +\frac14 \langle h_{i j,p }h_{k l,p} \rangle \right) u^i u^j u^k u^l \right]\, . 
\end{align}
We want to evaluate the above on  a thermal state of gravitons at temperature $T = 1/\beta$. 

The quantized graviton field can be expanded as
\begin{equation}
h_{ij}(\vec x) = \int (dp) \ \epsilon^\lambda_{ij}(\vec p)\left[\alpha_\lambda(p) e^{i \vec p \cdot \vec x} + \alpha^\dagger_\lambda(p) e^{-i \vec p \cdot \vec x}\right].
\end{equation}
In the above, 
\begin{equation}
(dp) \equiv \frac{d^3p}{(2 \pi)^3 2 p}, \qquad [\alpha_\lambda^\dagger(p), \alpha_{\lambda '}^\dagger(p')] = (2\pi)^3 (2  p) \delta^3(\vec p - \vec p\, ') \delta_{\lambda \lambda'}, 
\end{equation}
and 
\begin{equation}
\epsilon^+_{ij}(p^3) =  \begin{pmatrix}
0&0&0&0\\
0&1&0&0\\
0&0&-1&0\\
0&0&0&0
\end{pmatrix}, \qquad 
\epsilon^\times_{ij}(p^3) =  \begin{pmatrix}
0&0&0&0\\
0&0&1&0\\
0&1&0&0\\
0&0&0&0
\end{pmatrix}.
\end{equation}

It is useful to evaluate
\begin{equation} 
\sum_\lambda \epsilon^\lambda_{ij}(\vec p) \epsilon^\lambda_{kl}(\vec p)= D_{ik} D_{jl} + D_{il} D_{jk} - D_{ij} D_{kl},
\end{equation}
where 
\begin{equation}
D_{ij} \equiv \delta_{ij} - \frac{p_i p_j}{p^2}\, .
\end{equation}

Another basic ingredient for this calculation is the following 
\begin{equation}
\frac{Tr\left(\alpha^\dagger_\lambda (p') \alpha_{\lambda'}(p) e^{-\beta H}\right)}{Z} = \frac{(2\pi)^3 (2  p) \delta^3(\vec p - \vec p\, ') \delta_{\lambda \lambda'}}{e^{\beta p} -1}\, .
\end{equation}

One can then proceed to calculate the first expectation value in~\eqref{app-c1}. In order to subtract the divergent vacuum contribution all quadratic operators will be normal ordered. One finds 
\begin{equation}
\langle \dot h_{ij} \dot h_{kl} \rangle = 2 \int (dp) \sum_\lambda \epsilon^\lambda_{ij}(\vec p) \epsilon^\lambda_{kl}(\vec p) \frac{p^2}{e^{\beta p} -1}
\end{equation}
In order to evaluate this and some of the expression that will follow there is a basic integral to calculate once and for all, 
\begin{equation}
I \equiv \int (dp)  \frac{p^2}{e^{\beta p} -1} = \frac{\pi^2}{60 \beta^4}\, .
\end{equation}
Other integrals can be expressed in terms of $I$, 
\begin{align} 
\int (dp)  \frac{p_i p_j}{e^{\beta p} -1} & = \frac{I}{3} \delta_{ij}, \\
\int (dp)  \frac{p_i p_j p_k p_l}{p^2(e^{\beta p} -1)} & = \frac{I}{15} \delta_{ijkl}\\
\int (dp)  \frac{p_i p_j p_k p_lp_p p_q}{p^4(e^{\beta p} -1)} & = \frac{I}{105} \delta_{ijklpq},
\end{align}
where we have introduced some totally symmetrized delta functions 
\begin{align}
\delta_{ijkl} &= \delta_{ij} \delta_{kl} + \delta_{ik}  \delta_{jl} +  \delta_{il}  \delta_{jk} \\
\delta_{ijklpq} & = \delta_{ij} \delta_{kl} \delta_{pq} + 14 {\rm \ other \ terms}\, .
\end{align}
Note that, in the above formulas, an odd number of $p$s at the numerator would give zero. This is why the line~\eqref{app-c2} identically vanishes. As for the other pieces we basically need two expectation values that we compute in the following, 
\begin{align}
\langle \dot h_{ij} \dot h_{kl} \rangle &= \frac{4 I}{15} ( 3 \delta_{ik}\delta_{jl} + 3 \delta_{il}\delta_{jk}-2 \delta_{ij}\delta_{kl}),\\
\langle \dot h_{ij,p} \dot h_{kl,q} \rangle & = 2 I \left[\frac{\delta_{pq}}{3} (\delta_{ik} \delta_{jl} + \delta_{il} \delta_{jk} - \delta_{ij} \delta_{kl}) \right. \\
&\left. -\frac{1}{15}(\delta_{ik}\delta_{jlpq} + \delta_{jl}\delta_{ikpq} + \delta_{il}\delta_{jkpq} + \delta_{jk}\delta_{ilpq} - \delta_{ij}\delta_{klpq}  - \delta_{kl}\delta_{ijpq}  ) + \frac{1}{105} \delta_{ijklpq}\right]\, .\nonumber
\end{align}

With the last expression one can calculate the various contractions of~\eqref{app-c3},
\begin{equation}\nonumber
\langle h_{i p,j}h_{k p,l} \rangle u^i u^j u^k u^l = \frac45 I , \quad \langle h_{i p,j}h_{k l,p} \rangle  u^i u^j u^k u^l = \frac{4}{15} I, \quad  \langle h_{i j,p }h_{k l,p}\rangle u^i u^j u^k u^l  = \frac43 I
\end{equation}

The final result, introducing the canonical normalization for the graviton field, is
\begin{equation}
C = -\frac{49 \pi^2}{60^2} \frac{T^4}{M_P^2} t^4\, .
\end{equation}

\renewcommand{\baselinestretch}{1}\small
\bibliographystyle{ourbst}
%\bibliography{replicaBib}
\bibliography{references}
\end{document}